\documentclass[aps,pra,reprint,superscriptaddress,showpacs]{revtex4-1}
\usepackage[utf8]{inputenc}
\usepackage{graphicx}
\usepackage{amssymb,amsmath,amsthm,amsfonts}
\usepackage{hyperref}
\usepackage{bm}% bold math
\usepackage{subfloat}
\usepackage{float}
\usepackage{color}
\usepackage[caption=false,justification=justified]{subfig}
\usepackage{xcolor}
\usepackage[normalem]{ulem}

\newcommand\sbullet[1][.5]{\mathbin{\vcenter{\hbox{\scalebox{#1}{$\bullet$}}}}}

\newcommand{\nam}[1]{\textbf{\color{red}[Nam: #1]}}
\newcommand{\ming}[1]{\textbf{\color{blue}[Ming: #1]}}
\newcommand{\comment}[1]{\textbf{\color{red} }}

\begin{document}

\title{Generalized Hamiltonian to describe imperfections in ion-light interaction}
\author{Ming Li}
\affiliation{IonQ, College Park, MD 20740, USA}
\author{Kenneth Wright}
\affiliation{IonQ, College Park, MD 20740, USA}
\author{Neal C. Pisenti}
\affiliation{IonQ, College Park, MD 20740, USA}
\author{Kristin M. Beck}
\altaffiliation[Current address: ]{Physics Division, Physical and Life Sciences, Lawrence Livermore National Laboratory, Livermore, California 94550, USA}
\affiliation{IonQ, College Park, MD 20740, USA}
\author{Jason H. V. Nguyen}
\affiliation{IonQ, College Park, MD 20740, USA}
\author{Yunseong Nam}
\email{nam@ionq.co}
\affiliation{IonQ, College Park, MD 20740, USA}
\date{\today}

\begin{abstract}
We derive a general Hamiltonian that governs the interaction between an $N$-ion chain and an externally controlled laser field, where the ion motion is quantized and the laser field is considered beyond the plane-wave approximation. This general form not only explicitly includes terms that are used to drive ion-ion entanglement, but also a series of unwanted terms that can lead to quantum gate infidelity. We demonstrate the power of our expressivity of the general Hamiltonian by singling out the effect of axial mode heating and confirm this experimentally. 
%We discuss pathways forward in furthering the trapped-ion quantum computational quality that will help inform hardware design decisions.
We discuss pathways forward in furthering the trapped-ion quantum computational quality, guiding hardware design decisions.
\end{abstract}

%\pacs{}

\maketitle

%%%%%%%%%%%%%%%%%%%%%%%%%%%%%%%%%%%%%%%%%%%%%%%%%%%%%%%%%
\section{Introduction}

% \ming{Kristi: Suggestion: Make it clear that what these proposals share is limiting (usually to 2 ions) the ions that are involved in 2-qubit gates. This will more concretely link the demonstrations to the design and state the key system components (transport, limited connectivity) you contrast with the single chain architecture in the next paragraph.  Possible rewording: To this end, several pathways have been proposed and demonstrated where ion qubits are joined and separated during the quantum operations so that gates are only performed on short chains.}
Trapped ions represent a promising platform for universal quantum computation,
and high-fidelity quantum gates have already been demonstrated on short chains of one or two qubits~\cite{JGaebler16,CBallance16}.
%beating the error threshold required for fault tolerance \cite{}.
However, further improvements to gate fidelity and qubit count are necessary to bridge the gap between these %\old{early} 
academic %\old{systems} 
demonstrations and a commercially viable quantum computer.
%For the trapped-ion quantum computing platform to now go beyond being consigned to a pure academic interest and to start generating commercial value of an industrial scale, it becomes critical that the trapped-ion quantum computer further improves its quantum gate fidelity, while also increasing the number of qubits.
To this end, several pathways have been proposed and demonstrated~\cite{qccd,qccdHoneywell,photonicInterconnect},
where ion qubits are joined and separated during the quantum operations so that gates are only performed on short chains.
% where ion qubits are separated in space at a given time during a quantum program execution.
They however come at the cost of sparse qubit connectivity between the qubits, i.e., a direct implementation of qubit-to-qubit interaction between an arbitrary pair of qubits is impossible,
a known source of overhead in performing quantum computation \cite{Linke_2017,LowCost,IEEE}.
They also complicate the hardware design, making 
the quantum hardware more error prone.
% a streamlined and high-performance quantum program execution
%high-fidelity gate operation 
% more technically challenging. %interfering with the fidelity effort.
% \ming{Kristi: Can you be more specific with the 'more challenging'? }
%, including the quantum charge-coupled device (QCCD) architecture~\cite{qccd,qccdHoneywell}, where short chains of ion qubits are shuttled in and out of a designated inter-qubit interaction region or photonic %interconnects~\cite{photonicInterconnect}, where many ion-trap modules containing a long chain of ions are networked.
%These approaches however sacrifice the full connectivity between the ion qubits, a feature known to provide a decisive advantage in executing numerous quantum programs \cite{}.

%To thus retain the computational advantages provided by the full connectivity as much as possible, while inching toward a larger and high-performing trapped-ion quantum computer, 
To thus move towards a larger and high-performing trapped-ion quantum computer,
investigating in detail the mechanisms by which quantum computational errors may incur due to holding a modest-sized ion chain 
%that admits an entangling-gate implementation over an arbitrary pair of qubits 
becomes an important task. 
By successfully addressing the identified mechanisms, one can maximize the computational quality 
obtained from these scalable trapped-ion quantum computer architectures, while at the same time
also providing an additional advantage in quantum circuit implementation, should the addressing 
offer a way to reliably perform quantum computation over a longer chain of ions 
that admits an all-to-all qubit connectivity~\cite{Linke_2017,EASE}.
%efficiency provided by the large set of fully-connected ion qubits in each chain of ions in either the QCCD or photonic interconnect architectures. 
We aim to address this challenge by systematically expanding the light-matter interaction Hamiltonian, 
used to drive quantum gate operations on a trapped-ion quantum computer, in imperfections such as 
%\nam{list the expansion parameters here} 
misalignment, defocus, and ion motion.
%that arise primarily due to long-chain requirements. 
We focus in particular on the case where a two-photon Raman transition is used to implement a quantum logic gate. Our analytical results for the coupling of the internal qubit degrees of freedom to the quantized external motion of the ion chain accurately quantifies the role of ion-beam geometry in determining the quantum gate fidelity, in addition to the sensitivity of the fidelity with respect to ion-chain heating. We experimentally confirm the validity of our model and show compensating pulse sequences \cite{PulseSeqRMP,PulseSeq} can decisively help boost the fidelity of trapped-ion quantum computers. % with long ion chains.

In Sec.~\ref{sec:theory} we introduce
the governing effective Hamiltonian of a trapped-ion quantum computer, equipped
with two ion-addressing Raman beams. In Sec.~\ref{sec:Gaussian} we focus on
the Gaussian beam, a prototypical example used widely in the 
trapped-ion quantum computing community, and derive a suite of expressions
required for the generalized Hamiltonian. We then derive an approximate
Hamiltonian from the generalized Hamiltonian using realistic parameters
in Sec.~\ref{sec:error_analysis} and compare our theoretical results
with the experimental results in Sec.~\ref{sec:experiment}.
We discuss our work in Sec.~\ref{sec:discussion} and conclude our manuscript in Sec.~\ref{sec:conclusion}.

We note that related and similar results that mainly focus on the effect of ion motion 
perpendicular to the Gaussian beams used to drive quantum gates 
have recently been reported in \cite{marko,WesAxial}. %\nam{Marko paper and that other paper.}. 
% \nam{Emphasize the difference from these works here.}
Our work takes a more general approach in deriving the effective interaction
Hamiltonian associated with a spatially dependent light field, which enables 
systematic and quantitative error analysis for a variety of
hardware implementations of trapped-ion based quantum computers.
Our methodology further enables zooming in on individual error sources and
provide guidance in devising appropriate error mitigation strategies 
via the expressive power offered by our generalized Hamiltonian.
\section{General theory derivation}
\label{sec:theory}

%Full derivation assuming a generic spacial dependent electric field in all three dimensions. 

In this section, we derive the
Hamiltonian $H$ that couples the motional degrees of freedom of the
ions with their internal degrees of freedom through the spatial dependent Raman beams.
This Hamiltonian can be separated into two parts as
\begin{equation}
H = H_0 + H_I \;,
\end{equation}
where $H_0$ includes the internal and the motional degrees of freedom
for the ions independently and $H_I$ describes the light-matter 
interaction that couples the two.
%We consider the interaction between the Raman beams and the ions in a
%$N$-ion Coulomb crystal. 
Treating the laser field of the Raman beams
classically, while treating the rest of the system quantum mechanically, together with the dipole approximation, $H_I$ can simply be written as
\begin{equation}
H_I = - \sum_{k=1}^N \vec{E}\cdot\vec{d}_k \;,
\label{InteractionHam_original}
\end{equation}
where $\vec{E}$ is the electric field and
$\vec{d}_k$ is the dipole operator of the $k$th ion out of $N$ total number of ions.
%In a linear ion crystal, we assume the harmonic approximation 
%for the collective motion, and consider an effective
%two-level system for each ion as a qubit. 
For $H_0$ we consider ions that are confined in a linear Paul trap along the
potential-null line. We assume the harmonic approximation of the
collective motion of the ions. Focusing now on an effective two-level
system for the internal degrees of freedom of interest for each ion as a qubit,
we can then write
\begin{equation}
H_0 = \sum_{k=1}^{N}\frac{\hbar\omega_k^{\mathrm{qbt}}}{2}
    \hat{\sigma}_k^{z} + \sum_{p=1}^{3N}\hbar\omega_p
    \left(\hat{a}_p^\dag\hat{a}_p+\frac{1}{2}\right) \;,\label{eq:effectiveH0}
\end{equation}
where $\hbar$ is the reduced Planck constant, 
$\omega_k^{\mathrm{qbt}}$ is the effective qubit angular frequency 
for the $k$th ion, $\hat{\sigma}_k^\alpha$ with $\alpha = x$, $y$, or $z$ 
is the Pauli matrix along the $\alpha$-axis,
$\omega_p$ is the normal mode frequency of the $p$th normal mode with
Fock state creation and annihilation operators $\hat{a}_p^\dag$ and $\hat{a}_p$.
% \ming{Kristi: Suggestion: A cartoon of the axis definitions with an ion chain would make the intended geometry more accessible and would serve as a good reference for the experimental parameters later on. (x,y,z axes are not defined until the top of page 3) While it is not necessary for the derivation at this stage, I got distracted by the fact that they weren't yet defined. }

From here on, we consider only two Raman beams that drive
qubit transitions on a particular ion $k$, and we drop the ion
index wherever contextually clear for simplicity. The electric field near the ion is given by
\begin{equation}
\vec{E} = \vec{E}_1 + \vec{E}_2 \;,
\end{equation}
where the individual electric field $\vec{E}_b$ with
$b = 1$ or $2$ can be written as
\begin{equation}
\vec{E}_b = \hat{\epsilon}_b e^{i\omega_b t} E_b
  \left(\vec{r}_b\right) e^{i\Phi_b\left(\vec{r}_b\right)}
  +\mathrm{h.c.},
\end{equation}
where $\hat{\epsilon}_b$ is the polarization vector, $\omega_b$ is
the angular frequency, $E_b$ and $\Phi_b$ are
real functions of the ion position in each of the beam propagation
coordinates $\vec{r}_b$, and $\mathrm{h.c.}$ denotes 
the Hermitian conjugate. After 
%a standard maneuver of 
adiabatic elimination of the excited internal states of an ion \cite{EBrion2007},
% \nam{Is there a neat reference we can provide for this? Or should we show this explicitly in an appendix?}
we can approximate the individual summand $H_{I,k}$ of the interaction Hamiltonian $H_I$ in (\ref{InteractionHam_original}) 
as
\begin{equation}
H_{I,k} = \bar{D} e^{i(\omega^{\mathrm{qbt}}+\Delta\omega)t} E_1 e^{i\Phi_1} 
    E_2 e^{-i\Phi_2} \hat{\sigma}_k + \mathrm{h.c.} \;,
    \label{eq:HI}
\end{equation}
where $\bar{D}$ is an effective dipole constant, $\Delta\omega$
is the effective two-photon detuning of the Raman transition from the qubit
transition, and $\hat{\sigma}_k$ is a qubit spin operator which depends on the
details of the Raman transition scheme.
% \ming{Need to revisit. It's not necessary to have $\hat{\sigma^+_k}$,
% can be any (hermitian or non-hermitian) operator depending on
% the Raman transition scheme. Do we want to be general or stick to a
% specific scheme?}
Here, without loss of generality, we assume 
$\omega_1 - \omega_2 = \omega^{\mathrm{qbt}} + \Delta\omega$, 
in other words, the transition 
from qubit state
$|\hspace{-0.25em}\downarrow\rangle$ to 
$|\hspace{-0.25em}\uparrow\rangle$ requires 
a photon absorption from 
beam 1 and a photon emission from beam 2.

The coupling of the beam to the motional degrees of freedom of the ion
is embedded in the $\vec{r}_b$ dependent terms in (\ref{eq:HI}).
To rewrite them in terms of the normal mode operators $\hat{a}_p$
and $\hat{a}^\dag_p$, we first rewrite the ion position with respect
to each beam as
\begin{equation}
\vec{r}_b = \vec{r}_b^{(0)} + \sum_{\alpha_b} 
    \zeta_{\alpha_b} \hat{e}_{\alpha_b} \;,
\end{equation}
where $\vec{r}_b^{(0)}$ is the equilibrium position of the ion
in the beam coordinate $\{x_b, y_b, z_b\}$ and 
$\hat{e}_{\alpha_b}$ is the unit vector along the
direction of the axis $\alpha_b = x_b$, $y_b$, or $z_b$.
Then we can expand the terms $E_b$ and $e^{\pm i\Phi_b}$
near $\vec{r}_b^{(0)}$ with respect to $\zeta_{\alpha_b}$.
Lastly, we can quantize the ion motion from the equilibrium position and 
rewrite it in terms of the normal mode operators as
\begin{equation}
\hat{\zeta}_{\alpha_b}=\sum_{p=1}^{3N} \zeta_{p}^{(0)}
    \nu_p^{\alpha_b}\left(\hat{a}_p+\hat{a}_p^\dag\right) \;,
\label{eq:modes}
\end{equation}
where $\zeta_{p}^{(0)} = \sqrt{\hbar/2m\omega_p}$
is the spread of the zero-point wavefunction of mode $p$ with
the mass $m$ of the ion and $\nu_p^{\alpha_b}$ is a matrix element
of the inverse of the mode vector matrix~\cite{DFVJames1998}. %,JPHome2011}. 
% \nam{This reference is not actually the primary reference for this. I inserted the reference anyways but we may want to revisit.}
% \ming{D. F. V. James, Appl. Phys. B {\bf 66}, 181 (1998)?}
% \nam{Yes, please use this. Once addressed, please remove the comments.}
%\nam{Do we not need to say about where the origin of the beam-coordinate system is?}
%\ming{I don't think we need it here. I am adding a sentence in the next section for the Gaussian beam case.}
A full accounting of the ion-laser interaction can then be made by expanding the appropriate form of $E_be^{\pm i\Phi_b}$ for a given experimental context.

%%%%%%%%%%%%%%%%%%%%%%%%%%%%%%%%%%%%%%%%%%%%%%%%%%%%%%%%%
\section{Application to elliptical astigmatic Gaussian beams}
\label{sec:Gaussian}

In this section, we derive a convenient expression for the electric field that an illuminated ion sees, subject to unavoidable non-idealities that exist in a realistic ion-beam setup. Examples of non-idealities include beam
misalignment and defocus with respect to the equilibrium
position of the ion, unintended ion motion, etc.
Specifically, in Sec.~\ref{sec:beam}, we define useful notations for a Gaussian profile of a coherent beam that we then use in Sec.~\ref{sec:expansion} to derive a suite of series expansions that explicitly depend on the non-ideal parameters that comprise the electric field expression. In Sec.~\ref{sec:channel} we briefly discuss the ways in which noise may now couple into our system then lay out a strategy to use our derived expressions for an efficient and systematic error analysis in practice.

%%%%%%%%%%%%%%%%%%%%%%%%%%%%%%%%%%%%%
\subsection{Elliptical astigmatic Gaussian beam}
\label{sec:beam}

For the remainder of this paper, we focus on a particular form of $E_be^{\pm i\Phi_b}$ commonly used in experimental settings, an elliptical Gaussian beam with simple astigmatism~\cite{AESiegman1986}.
%We are interested in the spatial dependent terms of a particular
% Raman beam in the general form of $E_be^{\pm i\Phi_b}$.
% In particular, we consider in this subsection 
% the Raman beams that are simple (orthogonal) astigmatic
% fundamental Gaussian beams~\cite{AESiegman1986},
% the kind of beams that are most often used 
% in trapped-ion quantum computers \cite{}. %\nam{Recommended References?}
%\ming{For Ken and Neal: Do we need to say something about general astigmatism 
%(instead of the simple astigmatism case here), higher order Gaussian beams, and aberrations?}\neal{my feeling is we should make a more general statement about the approach, eg, doing an expansion about the zero-point of the oscillator wavefunction, and then work the specific example of elliptical gaussian beam as a ``very common example''; maybe reference Wes' paper as an example where higher order HG beams are used, and mention the existance of aberrations etc which, if characterized, could be thrown into this formalism to calculate the relevant couplings.}
The beam amplitude $E_b$ and phase angle $\Phi_b$ can be written %in their individual coordinate systems 
using the ion position $\vec{r}_b = x_b\hat{e}_{x_b}
+y_b\hat{e}_{y_b}+z_b\hat{e}_{z_b}$ as
\begin{align}
E_b(\vec{r}_b) &= \sqrt{\frac{P_b}{\pi w_{x_b}w_{z_b}}} e^{-\left(
    \frac{x_b^2}{w_{x_b}^2} + \frac{z_b^2}{w_{z_b}^2}\right)} \;, \nonumber \\
\Phi_b(\vec{r}_b) &= -k_b y_b + \eta_b - \frac{k_b}{2}\left(
    \frac{x_b^2}{R_{x_b}} + \frac{z_b^2}{R_{z_b}}\right) + \phi_b \;,
\label{eq:EPhi_pre_expansion}
\end{align}
where we assume the beam propagates along the $y_b$-axis and the
two principal axes are along the $x_b$- and the $z_b$-axes.
Here, $P_b$ is the power of the beam, $k_b = 2\pi/\lambda_b$ is the
wavevector with $\lambda_b$ the wavelength, and $\phi_b$ is a 
constant phase at the origin which can be chosen arbitrarily
along the $y_b$-axis. $w_{x_b}$ and $w_{z_b}$ are the
two principal semi axes of the spot ellipse at $y_b$, defined according to
\begin{equation}
w_{\alpha_b}(y_b) = w_{\alpha_b}^{\mathrm{f}}\sqrt{1+\left(
    \frac{y_b - y_{\alpha_b}^{\mathrm{f}}}
    {y^{\mathrm{R}}_{\alpha_b}}\right)^2} \;,
\label{eq:width_pre_expansion}
\end{equation}
where $w_{\alpha_b}^{\mathrm{f}}$ is the beam waist along the 
$\alpha_b$-axis at the focal point $y_{\alpha_b}^{\mathrm{f}}$ and
$y^{\mathrm{R}}_{\alpha_b}$ %\nam{think of a better notation..} 
%\ming{How about $y^{\mathrm{R}}_{\alpha_b}$?}
%\nam{Sure? Let's try it?}\ming{Replaced.}
is the Rayleigh range given by 
$\pi (w_{\alpha_b}^{\mathrm{f}})^2 / \lambda_b$.
The radii of curvature $R_{\alpha_b}$ are given by
\begin{equation}
R_{\alpha_b}(y_b) = \frac{\left(y_b - y_{\alpha_b}^{\mathrm{f}}\right)^2
    + {y^{\mathrm{R}}_{\alpha_b}}^2}{y_b - y_{\alpha_b}^{\mathrm{f}}} \;.
\label{eq:radii_pre_expansion}
\end{equation}
$\eta_b$ is the Gouy phase, i.e.,
\begin{equation}
\eta_b(y_b) = \frac{1}{2}\left(\arctan\frac{y_b - y_{x_b}^{\mathrm{f}}}
    {y^{\mathrm{R}}_{x_b}} + \arctan\frac{y_b - y_{z_b}^{\mathrm{f}}}
    {y^{\mathrm{R}}_{z_b}}\right) \;.
\label{eq:gouy_pre_expansion}
\end{equation}

\subsection{Expansion of the electric field}
\label{sec:expansion}

The spatially dependent terms $E_be^{\pm i\Phi_b}$ can be expanded from the focal points of the beam.
Denoting the ion equilibrium position as $\vec{r}_b^{(0)} = x_b^{(0)}\hat{e}_{x_b}
+y_b^{(0)}\hat{e}_{y_b}+z_b^{(0)}\hat{e}_{z_b}$ and the
ion excursion as $\{\zeta_{x_b},\zeta_{y_b},\zeta_{z_b}\}$,
we define the $y$-distance between the ion equilibrium position 
and the $x$ and $z$ focal points as
$y_{\alpha_b}^{(0)\mathrm{f}}\equiv y_b^{(0)}-y_{\alpha_b}^{\mathrm{f}}$.
Then, the regime we consider here, i.e.,
the ion does not venture outside of the Rayleigh range from 
each focal point of the corresponding principal axis, may be succinctly written as
$\left|y_{\alpha_b}^{(0)\mathrm{f}}\right|\ll y^{\mathrm{R}}_{\alpha_b}$.
%Similar derivation can be made outside of the Rayleigh range.
%When $\left|y_{\alpha_b}^{(0)\mathrm{f}}\right|\sim y_{\mathrm{R}_{\alpha_b}}$
%other treatments will be needed.
%We can independently Taylor expand the $E_b$ part and 
%the $e^{\pm i\Phi_b}$ part. 
%To better understand the structure of the expansion,
In this regime, %instead of directly Taylor expanding near the equilibrium position
%of the ion, we opt to expand the focal points of the two principal axes as
expanding $E_b$ and $e^{i\Phi_b}$ in (\ref{eq:EPhi_pre_expansion})
about the focal points of the two principal axes $y_{x_b}^f$ and $y_{z_b}^f$, 
together with (\ref{eq:width_pre_expansion}), (\ref{eq:radii_pre_expansion}), and (\ref{eq:gouy_pre_expansion}), 
we obtain
% \begin{widetext}
\begin{align}
E_b 
%     =& \; \sqrt{\frac{P}{\pi w_{x_b}^{\mathrm{f}}
%     w_{z_b}^{\mathrm{f}}}}
%     \left[\sum_{n=0}^\infty (-1)^n \frac{(4n-3)!!!!}{(4n)!!!!}
%       \left(\frac{\zeta_{y_b}+y^{(0)\mathrm{f}}_{x_b}}{y_{\mathrm{R}_{x_b}}}
%       \right)^{2n}\right]
%     \left[\sum_{n=0}^\infty (-1)^n \frac{(4n-3)!!!!}{(4n)!!!!}
%       \left(\frac{\zeta_{y_b}+y^{(0)\mathrm{f}}_{z_b}}{y_{\mathrm{R}_{z_b}}}
%       \right)^{2n}\right] \nonumber\\
%     & \left[\sum_{n=0}^\infty\sum_{m=0}^\infty\frac{(-1)^{n+m}(n+m-1)!}
%     {(n-1)!\,n!\,m!}\left(\frac{\zeta_{x_b}+x^{(0)}_b}{w_{x_b}^{\mathrm{f}}}\right)^{2n}
%     \left(\frac{\zeta_{y_b}+y^{(0)\mathrm{f}}_{x_b}}{y_{\mathrm{R}_{x_b}}}
%       \right)^{2m}\right] \nonumber\\
%     & \left[\sum_{n=0}^\infty\sum_{m=0}^\infty\frac{(-1)^{n+m}(n+m-1)!}
%     {(n-1)!\,n!\,m!}\left(\frac{\zeta_{z_b}+z^{(0)}_b}{w_{z_b}^{\mathrm{f}}}\right)^{2n}
%     \left(\frac{\zeta_{y_b}+y^{(0)\mathrm{f}}_{z_b}}{y_{\mathrm{R}_{z_b}}}
%       \right)^{2m}\right] 
%   \\
    \equiv& \; 
    \sqrt{\frac{P_b}{\pi w_{x_b}^{\mathrm{f}}w_{z_b}^{\mathrm{f}}}} \, 
    A_1(\lambda_{x_b}^{(0)},\hat{\lambda}_{x_b}) \,
    A_1(\lambda_{z_b}^{(0)},\hat{\lambda}_{z_b}) \, \nonumber\\
    & %\hspace{-1.5em}
    A_2(\lambda_{x_b}^{(0)},\hat{\lambda}_{x_b},
        \gamma_{x_b}^{(0)},\hat{\gamma}_{x_b}) \, 
    A_2(\lambda_{z_b}^{(0)},\hat{\lambda}_{z_b},
        \gamma_{z_b}^{(0)},\hat{\gamma}_{z_b}) \;
    \label{eq:Eb}
\end{align}
and
\begin{align}
e^{\pm i\Phi_b} 
%     =& \; e^{\pm i(\phi_b - k_by_b^{(0)})}
%     \left[\sum_{n=0}^\infty\frac{(\mp i)^n}{n!}(k_b \zeta_{y_b})^n\right] \nonumber\\
%     & \left[\sum_{n=0}^\infty\frac{(\pm i)^n}{2^n\,n!}\left\{
%       \sum_{m=0}^\infty\frac{(-1)^m}{2m+1}\left(
%       \frac{\zeta_{y_b}+y_{x_b}^{(0)\mathrm{f}}}{y_{\mathrm{R}_{x_b}}}
%       \right)^{2m+1}
%     \right\}^n\right] 
%     \left[\sum_{n=0}^\infty\frac{(\pm i)^n}{2^n\,n!}\left\{
%       \sum_{m=0}^\infty\frac{(-1)^m}{2m+1}\left(
%       \frac{\zeta_{y_b}+y_{z_b}^{(0)\mathrm{f}}}{y_{\mathrm{R}_{z_b}}}
%       \right)^{2m+1}
%     \right\}^n \right]\nonumber\\
%     & \left[\sum_{n=0}^\infty\sum_{m=0}^\infty
%     \frac{(\pm i)^n (-1)^{(n+m)} (n+m-1)!}{(n-1)!\,n!\,m!}
%     \left(\frac{\zeta_{x_b}+x^{(0)}_b}{w_{x_b}^{\mathrm{f}}}\right)^{2n}
%     \left(\frac{\zeta_{y_b}+y_{x_b}^{(0)\mathrm{f}}}{y_{\mathrm{R}_{x_b}}}
%       \right)^{2m+n}\right] \nonumber\\
%     & \left[\sum_{n=0}^\infty\sum_{m=0}^\infty
%     \frac{(\pm i)^n (-1)^{(n+m)} (n+m-1)!}{(n-1)!\,n!\,m!}
%     \left(\frac{\zeta_{z_b}+z^{(0)}_b}{w_{z_b}^{\mathrm{f}}}\right)^{2n}
%     \left(\frac{\zeta_{y_b}+y_{z_b}^{(0)\mathrm{f}}}{y_{\mathrm{R}_{z_b}}}
%       \right)^{2m+n}\right] 
%   \\
    \equiv& \; e^{\pm i(\phi_b - k_by_b^{(0)})} 
    B_0^{\pm}(\hat{\beta}_b) \,
        %\nonumber\\
    B_1^{\pm}(\lambda_{x_b}^{(0)},\hat{\lambda}_{x_b}) \,
    B_1^{\pm}(\lambda_{z_b}^{(0)},\hat{\lambda}_{z_b}) \, \nonumber\\
    &
    B_2^{\pm}(\lambda_{x_b}^{(0)},\hat{\lambda}_{x_b},
        \gamma_{x_b}^{(0)},\hat{\gamma}_{x_b})
    B_2^{\pm}(\lambda_{z_b}^{(0)},\hat{\lambda}_{z_b},
        \gamma_{z_b}^{(0)},\hat{\gamma}_{z_b}) \;,
     \label{eq:Phib}
% Another form  
% e^{\pm i\Phi_b} &= e^{\pm i(\phi_b - k_by_b^{(0)})}
%     \sum_{n=0}^\infty\frac{(\mp i)^n}{n!}(k_b \zeta_{y_b})^n \nonumber\\
%     & \sum_{n=0}^\infty\frac{(\pm i)^n}{2^n\,n!}\left[
%       \sum_{m=0}^\infty\frac{(-1)^m}{2m+1}\left(
%       \frac{\zeta_{y_b}+y_{x_b}^{(0)\mathrm{f}}}{y_{\mathrm{R}_{x_b}}}
%       \right)^{2m+1}
%     \right]^n 
%      \sum_{n=0}^\infty\frac{(\mp i)^n}{n!}
%     \left(\frac{\zeta_{x_b}+x^{(0)}_b}{w_{x_b}^{\mathrm{f}}}\right)^{2n}\left[
%       \sum_{m=0}^\infty (-1)^m\left(
%       \frac{\zeta_{y_b}+y_{x_b}^{(0)\mathrm{f}}}{y_{\mathrm{R}_{x_b}}}
%       \right)^{2m+1}
%     \right]^n \nonumber\\
%     &\sum_{n=0}^\infty\frac{(\pm i)^n}{2^n\,n!}\left[
%       \sum_{m=0}^\infty\frac{(-1)^m}{2m+1}\left(
%       \frac{\zeta_{y_b}+y_{z_b}^{(0)\mathrm{f}}}{y_{\mathrm{R}_{z_b}}}
%       \right)^{2m+1}
%     \right]^n
%     \sum_{n=0}^\infty\frac{(\mp i)^n}{n!}
%     \left(\frac{\zeta_{z_b}+z^{(0)}_b}{w_{z_b}^{\mathrm{f}}}\right)^{2n}\left[
%       \sum_{m=0}^\infty (-1)^m\left(
%       \frac{\zeta_{y_b}+y_{z_b}^{(0)\mathrm{f}}}{y_{\mathrm{R}_{z_b}}}
%       \right)^{2m+1}
%     \right]^n .
\end{align}
% \end{widetext}
where we split the spatial dependent terms into the $A$ and $B$ 
functions, defined according to 
\begin{widetext}
\begin{align}
    A_1(p_0,\hat{p}_1) &= \left[1 + (p_0+\hat{p}_1)^2\right]^{-1/4} =
      \sum_{l_p=0}^{\infty} \hat{p}_1^{l_p}
      \sum_{n=\lceil l_p/2 \rceil}^{\infty}
      (-1)^n \frac{(4n-3)!!!!}{(4n)!!!!} \binom{2n}{l_p}
      p_0^{2n-l_p} \;, \nonumber \\
%%%%%%%%%%%%%%%%%%%%%%%%%%%%%%%%%%%%%%%
    A_2(p_0,\hat{p}_1,q_0,\hat{q}_1) &= 
      \exp{\left[-\frac{(q_0+\hat{q}_1)^2}{1+(p_0+\hat{p}_1)^2}\right]} \nonumber \\
      &=
    %   \hspace{-0.5em} 
      \sum_{l_q,l_p=0}^{\infty}\hat{q}_1^{l_q}\hat{p}_1^{l_p}
    %   \hspace{-0.5em}
      \sum_{\substack{n=\lceil l_q/2 \rceil 
          \\ m=\lceil l_p/2 \rceil}}^{\infty}
    %   \hspace{-0.5em}
      \frac{(-1)^{n+m}}
      {n!}\binom{n+m-1}{m}\binom{2n}{l_q}\binom{2m}{l_p}
      q_0^{2n-l_q} p_0^{2m-l_p} \;, \nonumber \\
%%%%%%%%%%%%%%%%%%%%%%%%%%%%%%%%%%%%%%%
    B_0^{\pm}(\hat{p}_1) &= e^{\mp i \hat{p}_1} =
      \sum_{n=0}^\infty\frac{(\mp i)^n}{n!}\hat{p}_1^n \;, \nonumber\\
%%%%%%%%%%%%%%%%%%%%%%%%%%%%%%%%%%%%%%%
    B_1^{\pm}(p_0,\hat{p}_1) &= 
      \exp{\left[\frac{\pm i}{2}\arctan(p_0+\hat{p}_1)\right]} =
      \sum_{n=0}^\infty\frac{(\pm i)^n}{2^n\,n!}\left\{
      \sum_{l_p=0}^{\infty}\hat{p}_1^{l_p}
      \sum_{m=\lceil (l_p-1)/2 \rceil}^\infty
      \frac{(-1)^m}{2m+1}
      \binom{2m+1}{l_p}p_0^{2m+1-l_p} \right\}^n \;, \nonumber \\
%%%%%%%%%%%%%%%%%%%%%%%%%%%%%%%%%%%%%%%
    % B_2^{\pm}(p_0,\hat{p}_1,q_0,\hat{q}_1) &= 
    %   \sum_{l_q,l_p=0}^{\infty}\hat{q}_1^{l_q}\hat{p}_1^{l_p}
    %   \sum_{\substack{n,m \ge 0,\\2n\ge l_q,\\
    %         2m+n\ge l_p}}^{\infty}
    %   \frac{(\pm i)^n(-1)^{n+m}}
    %   {n!}\binom{n+m-1}{m}\binom{2n}{l_q}\binom{2m+n}{l_p}
    %   q_0^{2n-l_q} p_0^{2m+n-l_p} \;,
    B_2^{\pm}(p_0,\hat{p}_1,q_0,\hat{q}_1) &= 
      \exp{\left[\mp i\frac{(p_0+\hat{p}_1)(q_0+\hat{q}_1)^2}
      {1+(p_0+\hat{p}_1)^2}\right]} \nonumber \\
      &=
    %   \hspace{-0.5em} 
      \sum_{l_q,l_p=0}^{\infty}
      \hat{q}_1^{l_q}\hat{p}_1^{l_p}
    %   \hspace{-0.75em}
      \sum_{\substack{n=\lceil l_q/2 \rceil \\ 
           m=\lceil l_p/2 \rceil}}^{\infty}
    %   \hspace{-0.75em}
      \frac{(\mp i)^n(-1)^{m}}
      {n!}\binom{n+m-1}{m}\binom{2n}{l_q}\binom{2m}{l_p}
      (p_0+\hat{p}_1)^n q_0^{2n-l_q} p_0^{2m-l_p} \;, \nonumber \\
\label{eq:ABfunc}
\end{align}
\end{widetext}
where $\binom{\;\sbullet[0.4]\;}{\;\sbullet[0.4]\;}$ denotes a 
binomial coefficient, $\lceil{\,\sbullet[0.4]\,}\rceil$ denotes 
the ceiling function, $p_0$ may be $\lambda_{x_b}^{(0)}$ or 
$\lambda_{z_b}^{(0)}$, $\hat{p}_1$ may be $\hat{\lambda}_{x_b}$,
$\hat{\lambda}_{z_b}$, or $\hat{\beta}_b$, $q_0$ may be
$\gamma_{x_b}^{(0)}$ or $\gamma_{z_b}^{(0)}$,
$\hat{q}_1$ may be $\hat{\gamma}_{x_b}$ or $\hat{\gamma}_{z_b}$,
and
\begin{align}
    \hat{\beta}_b =&\; k_b\hat{\zeta}_{y_b} 
      =\sum_p c_{p,y_b}^{\beta} (\hat{a}_p+\hat{a}_p^\dag)\;,
        \nonumber\\
    \gamma_{x_b}^{(0)} = \frac{x^{(0)}_b}{w_{x_b}^{\mathrm{f}}} \;&,\;
    \hat{\gamma}_{x_b} = \frac{\hat{\zeta}_{x_b}}{w_{x_b}^{\mathrm{f}}}
      =\sum_p c_{p,x_b}^{\gamma} (\hat{a}_p+\hat{a}_p^\dag)\;,
        \nonumber\\
    \gamma_{z_b}^{(0)} = \frac{z^{(0)}_b}{w_{z_b}^{\mathrm{f}}} \;&,\;
    \hat{\gamma}_{z_b} = \frac{\hat{\zeta}_{z_b}}{w_{z_b}^{\mathrm{f}}}
      =\sum_p c_{p,z_b}^{\gamma} (\hat{a}_p+\hat{a}_p^\dag)\;,
        \nonumber\\
    \lambda_{x_b}^{(0)} = \frac{y_{x_b}^{(0)\mathrm{f}}}
        {y^{\mathrm{R}}_{x_b}} \;&,\;
    \hat{\lambda}_{x_b} = \frac{\hat{\zeta}_{y_b}}{y^{\mathrm{R}}_{x_b}}
      =\sum_p c_{p,x_b}^{\lambda} (\hat{a}_p+\hat{a}_p^\dag)\;,
        \nonumber\\
    \lambda_{z_b}^{(0)} = \frac{y_{z_b}^{(0)\mathrm{f}}}
        {y^{\mathrm{R}}_{z_b}} \;&,\;
    \hat{\lambda}_{z_b} = \frac{\hat{\zeta}_{y_b}}{y^{\mathrm{R}}_{z_b}}
      =\sum_p c_{p,z_b}^{\lambda} (\hat{a}_p+\hat{a}_p^\dag)\;
\label{eq:dimensionless}
\end{align}
are dimensionless. We quantize $\beta_b$, $\gamma_{\alpha_b}$, and
$\lambda_{\alpha_b}$ and use Eq.~(\ref{eq:modes})
to express the $A$ and $B$ functions defined in (\ref{eq:ABfunc}) 
in terms of normal mode ladder operators and collect all the
non-operator coefficients into the $c$-coefficients. Note that
the the $c$-coefficients are inversely proportional to the square-root
of their corresponding mode frequencies.
% We try to express the $A$ and $B$ functions in ascending order of 
% $\hat{p}_1 = \hat{\lambda}_{\alpha_b}$ or $\hat{\beta}_b$ and
% $\hat{q}_1 = \hat{\gamma}_{\alpha_b}$ whenever possible without
% compromising convergence of further analysis, 
% which corresponds to the increasing power in the ladder operators.
Further note that the $B_0^{\pm}$ terms
are conventionally used to formulate two-qubit entangling
gates, for instance the M{\o}lmer–S{\o}rensen protocol~\cite{MS1,MS2} 
or the Cirac-Zoller protocol~\cite{CZ}. It is sometimes easier
to maintain the exponential form for $B_0^{\pm}$ since its
exponent only has first order terms of ladder operators.

We briefly emphasize that the Hamiltonian framework detailed in this section is entirely general. It enables quantum hardware designers to straightforwardly assess the impact of a variety of experimental imperfections on the quantum computational fidelity. Our framework serves as a diagnostic tool that aids the designers locate the major sources of quantum computational errors, critical for developing powerful quantum computers.

\subsection{Error channels and analysis strategy}
\label{sec:channel}

By examining the expansions reported above, 
one can identify four general mechanisms by which 
a quantum computational error can occur through the 
spatially dependent terms. Firstly, any misalignment, defocus, or
distortion of a Raman beam that is not accounted for can lead to
quantum computational errors. 
Secondly, when there is stray field near the ion that are 
not compensated or accounted for, errors may also arise. 
In these two situations, errors propagate through all of the parameters
in (\ref{eq:dimensionless}). Specifically, they affect 
the scaled-position parameters $\gamma_{\alpha_b}^{(0)}$,
$\lambda_{\alpha_b}^{(0)}$, or the non-operator terms in the
definition of $\hat{\beta}_b$, $\hat{\gamma}_{\alpha_b}$, or 
$\hat{\lambda}_{\alpha_b}$, {\it i.e.} the zero-point spread $\zeta_p^{(0)}$, 
the matrix element $\nu_p^{\alpha_b}$ of the inverse mode vector matrix, 
the beam waist $w_{\alpha_b}^{\mathrm{f}}$,
or the Rayleigh range $y_{R_{\alpha_b}}$.
Thirdly, an error can occur through the so-called 
resonant terms that do not change the motional space, 
${\it i.e.}$ they have equal numbers of $\hat{a}_p$ and 
$\hat{a}_p^{\dag}$ operators. Any even total power
of $\hat{\beta}_b$, $\hat{\gamma}_{\alpha_b}$, and/or 
$\hat{\lambda}_{\alpha_b}$ 
would contain resonant terms. Apart from the trivial case of 
a constant term, all the other terms depend on the occupation 
of the motional Fock space. Thus any imprecise control or erroneous
information on the motional space can lead to quantum computational
errors in the manipulation of ion qubits. 
A classic example of this mechanism is 
the well-known Debye-Waller effect~\cite{NISTBible}. 
Lastly, the rest of the $\hat{a}_p$ and $\hat{a}_p^{\dag}$ 
dependent terms lead to an excursion in the phase space of 
the ion during a quantum gate operation. 
When such an excursion occurs and the ion is not returned 
to its initial position in the phase space after the completion
of the gate operation, it can lead to unwanted, lingering 
entanglement between the qubit space and the motional space, 
which are nontrivial to correct for. 
This effect could in part be suppressed by 
reducing the corresponding coefficient for the 
$\hat{a}_p$ and $\hat{a}_p^{\dag}$ dependent terms,
sufficient detunings from any motional sideband resonances, %\cite{},
or by actively shaping the gate pulse \cite{AMFM}.

% To further analyze the impact of undesirable couplings in $H_I$, 
% one could rearrange the terms in the summations in (\ref{eq:ABfunc})
% so that they reflect the increasing power in the operators
% $\hat{\gamma}_{\alpha_b}$ and $\hat{\lambda}_{\alpha_b}$ which correspond
% to power in the ladder operators. We present explicit formulae
% up to the second order in $\hat{p}_1$ or $\hat{q}_1$ for $A_{1,2}$ and 
% $B_{1,2}^{\pm}$ in the rest of this section 
% ($B_0^{\pm}$ already has the right form).
%\nam{Ming, is this what you meant?} {yes! thanks!}
%\nam{This paragraph needs work as per Ming's request -- will come back to it.}
%\ming{Tried to rework this paragraph here.}
%\ming{@Nam: please check again.}
In practice, it is cumbersome to directly
use the expressions of $A$ and $B$ functions in (\ref{eq:ABfunc}). 
A proper and justifiable truncation of the power series in (\ref{eq:ABfunc}) 
becomes an important task for an approximate yet effective error analysis.
We observe that each function in (\ref{eq:ABfunc}) 
can be written in the form of a summation of operators, 
$\sum_{ij} \hat{O}_{ij}$, where each operator is in the form of 
$c_{ij} \hat{p}_1^i\hat{q}_1^j$.
$c_{ij}$ here is a complex constant and $c_{00}$ is always non-zero.
Our task then boils down to neglecting some of the operators 
if their contribution to the Hamiltonian is small. 
To quantify the contribution, we  
use the operator norm $\|\hat{O}_{ij}\|$. 
We evaluate the norm in a large but finite
motional space, truncated such that 
realistic motional-space dynamics 
can be adequately captured within.
In the next section, we will perform
the error analysis for a realistic situation
and provide a concrete example.
We note in passing that 
we can rewrite the power series of the non-operator terms in 
(\ref{eq:ABfunc}) in a more compact way by examining terms 
with ascending power of $\hat{p}_1$. 
Doing so renders evaluating the size of the individual coefficients
of the powers of $\hat{p}_1$ more straightforward. 
We report the results for the first three orders in 
Appendix~\ref{sec:app_p1} for the convenience of the readers. 

% one may analyze the contribution
% of each term in the $A$ and $B$ functions in terms of its
% non-operator coefficients as well as the expectation value \nam{exp. value of? operator terms?},
% assuming a certain Fock state occupation of the motional space.
% One can then truncate the power series according to a 
% predetermined accuracy requirement. In the rest of this section,
% to help analyze the contribution of these terms, 
% we present the $A$ and $B$ functions in a way that sums up 
% the non-operator terms, {\it i.e.} terms of $p_0$ and $q_0$ 
% in (\ref{eq:ABfunc}). Such a compact form is possible
% when terms are collected with a specific given order of $\hat{p}_1$,
% which we show up to the second order. 
% \ming{Need help to reword a bit. Strictly speaking the expression for
% $B_2^{\pm}$ is not ordered by ascending power of $\hat{p}_1$ because
% of the $\hat{s}_\pm$ terms. This is not avoidable due to the
% form of the functions. Although there is explicit $\hat{p}_1$ terms
% up to the second order in the square bracket, and is what's used later
% for order of magnitude evaluation and truncation of the series.
% }

%\section{Error analysis for undesirable axial coupling in co-propagating Raman beam set-ups}
\section{Parallel Raman beam geometry}
\label{sec:error_analysis}

%\nam{The goal of this section is to compute $P_{\uparrow}$, an experimental quantity we can measure. This section consists largely of three parts: Parameter size introduction, Hamiltonian approximation, single-qubit gate unitary and its application to an initialized state to obtain $P_{\uparrow}$.}

% To illustrate how errors propagate, we investigate two 
% realistic Raman beam set-ups that are commonly used in 
% trapped-ion based quantum computers.
% We call them the co-propagating and the counter propagating
% schemes~\cite{ShantanuThesis}, which, as their names suggest, 
% one has the two Raman beams pointing along the same direction, 
% and the other has them pointing along opposite directions.
In this section, we provide a concrete analysis
based on (\ref{eq:ABfunc}) for a realistic set of 
Raman beam parameters relevant to contemporary trapped-ion quantum computing architectures
with parallel Raman beams configured to be either co- or counter- propagating.
In Sec.~\ref{sec:spec} we specify sizes of the parameters
commensurate to a contemporary trapped-ion quantum computer.
%\ming{The above sentence: the sizes is for our quantum computer and
%not general enough to cover others. Do we want to readjust this sentence?}
%\nam{Adjusted.}
In Sec.~\ref{sec:approx}, we follow through the error analysis
strategy laid out in Sec.~\ref{sec:channel} and present a simplified
version of the evolution operator that approximately describes 
the quantum state evolution. % to within $10^{-2}$ accuracy.
In Sec.~\ref{sec:Axial}, we subject our approximate evolution operator to application
and show in particular the significance of the axial mode temperature 
in determining quantum gate fidelity when using tightly focused Raman beams.
%, which we confirm experimentally in Sec.~\ref{sec:experiment}. 
%Similar experimental results have recently been reported in Ref.~\cite{marko}.
%\nam{The Marko paper discussion should go somewhere else?}

%\ming{@Nam: Here's a paragraph about intuitively why axial mode temperature
%decoheres gates. I added a similar paragraph under Eq. 25, but I
%think it might be good to emphasize it somewhere else too.
%It doesn't fit here, but I'm just leaving it here for now until 
%we find a better place for it..}
%\ming{I don't think we have mentioned axial temperature yet
%by now. I feel like the segue seems a bit abrupt.}

We emphasize that, intuitively speaking, the impact on quantum gate fidelity due excessive axial mode temperature is similar to that due to the Debye-Waller effect, {\it i.e.} the
%We emphasize that, intuitively speaking, 
%the way excessive axial mode temperature 
%affect quantum gate fidelity is similar to the way Debye-Waller
%effect impact quantum gate fidelity, {\it i.e.} the
Rabi frequency for driving the spin degree of freedom depends on
the phonon number of the axial motional state. 
Therefore, any distribution of motional state with a non-zero width 
directly translates to a distribution in the Rabi frequency
with a corresponding non-zero width that decoheres the quantum
gate operation.
We briefly re-discuss this point towards the end of Sec.~\ref{sec:Axial}
once we derive all necessary expressions for computing the quantum gate
fidelity.

\subsection{Parameter specifications}
\label{sec:spec}

We assume a linear chain of ions, addressed by an array of Raman
beams propagating in parallel,
capable of driving transitions between $|\hspace{-0.25em}\downarrow\rangle$
and $|\hspace{-0.25em}\uparrow\rangle$, tightly focused along 
the chain axis, to achieve individual addressability of qubits along the chain.
We refer to the normal modes of the ion chain depending on the dominant projection of their mode vector -- axial modes are predominantly along $\hat x_b$, horizontal modes along $\hat y_b$, and vertical modes along $\hat z_b$. 
Our coordinate systems are defined with respect to the axes of the Raman beams, which are assumed to propagate along $\hat y_b$ transverse to the chain axis, and exhibit an elliptical Gaussian profile with the loose (tight) dimension along $\hat z_b$ ($\hat x_b$).
The equilibrium position of each ion is assumed to reside near the focal point of each Raman beam, such that we satisfy $\gamma_{\alpha_b}^{(0)}, \lambda_{\alpha_b}^{(0)} \ll 1$ 
%and $\lambda_{\alpha_b}^{(0)} \ll 1$.

For a quantitative analysis, we consider Raman beams with a wavelength $\lambda = 355$~nm and a waist $w^{\rm f}_{x_b}$ ($w^{\rm f}_{z_b}$) larger or similar to $\sim 1~\mu$m ($\sim 5~\mu$m), as might be found on a $^{171}$Yb$^+$ trapped ion quantum computer.
% \ming{Need to revisit and update these numbers.}
% \old{The Rayleigh ranges for the two principal axes are then given by $y_{R_{x_b}}\gtrsim 10~\mu$m and $y_{R_{z_b}}\gtrsim 200~\mu$m.}
The corresponding Rayleigh range $y_{R_{x_b}}$ ($y_{R_{z_b}}$) is then larger or similar to $\sim 10~\mu$m ($\sim 200~\mu$m).
We assume the alignment errors in $|x_b^{(0)}|$ and $|z_b^{(0)}|$ are less than $100$~nm, and the focusing error in $|y_{\alpha_b}^{(0)\mathrm{f}}|$ is bounded by $10\%$ of the corresponding Rayleigh range. 
Then, we have $|\gamma_{x_b}^{(0)}|<0.1$, $|\gamma_{z_b}^{(0)}|<0.02$, $|\lambda_{x_b}^{(0)}|<0.1$, and $|\lambda_{z_b}^{(0)}|<0.1$.
Similarly, we quantify the alignment of a given normal mode to the dominant principle axes of the Raman beams by an error parameter $\varepsilon$, defined as the size of the maximal relative excursion of the mode vector matrix elements $\nu_p^{\alpha_b}$ in non-dominant principle axis directions scaled by a factor of $\sqrt{N}$. 
The magnitude of unintended projection of each mode vector 
along $\{\hat x, \hat y, \hat z\}$ are then bounded as given in Table~\ref{tb:modevec}.
For the system we consider, we assume $\varepsilon$ to be smaller than $0.05$.
Normal mode frequencies $\omega_p/2\pi$ are taken to be approximately 3~MHz horizontally, 2.5~MHz vertically, and depending on the number of ions, chain spacing, and the DC potential, anywhere from 150~kHz to $\sim 2$~MHz axially.
The resulting magnitude of the $c$-coefficients that appear in
(\ref{eq:dimensionless}) are summarized in Table~\ref{tb:ccoeffs}.

We utilize Doppler cooling on the $^2S_{1/2}$ to $^2P_{1/2}$
transition to cool the Yb$^+$ ions. The mode temperatures after the
Doppler cooling are given by the average phonon number at the 
Doppler limit $\bar{n}^D_p = \Gamma/2\omega_p$ where 
$\Gamma = 2\pi\times19.6~$MHz
is the natural linewidth of the excited $^2P_{1/2}$ state.
Thus the average phonon number at the Doppler limit is $\sim 4$ quanta
for the non-axial modes, and can range from $\sim 5$ to $\sim 70$
quanta for the axial modes depending on the mode frequencies.
For the horizontal modes, in the case of counter-propagating
set up, we apply a sideband cooling sequence, which consists of coherent
red sideband pulses, followed by optical pumping. This consistently
cools the horizontal modes to $\bar{n} \lesssim 0.1$.

\begin{table}
    \centering
    \caption{ Alignment of the principal axes of the Gaussian beams
        with the mode vectors of different sets of modes.
        \label{tb:modevec}}
    \begin{ruledtabular}
    \begin{tabular}{cccc}
    &  Axial modes &  Horizontal modes  &  Vertical modes \\
    \hline
    $|\nu_p^{x_b}|$ & $\sim 1/\sqrt{N}$ & $< \varepsilon/\sqrt{N}$ & $< \varepsilon/\sqrt{N}$ \\
    $|\nu_p^{y_b}|$ & $< \varepsilon/\sqrt{N}$ & $\sim 1/\sqrt{N}$ & $< \varepsilon/\sqrt{N}$ \\
    $|\nu_p^{z_b}|$ & $< \varepsilon/\sqrt{N}$ & $< \varepsilon/\sqrt{N}$ & $\sim 1/\sqrt{N}$
    \end{tabular}
    \end{ruledtabular}
\end{table}

\begin{table*}
    \centering
    \caption{ Estimates of the magnitude of the $c$-coefficients
    in Eq.~(\ref{eq:dimensionless}) using the realistic experimental
    parameters and conditions detailed in the main text of Sec.~\ref{sec:error_analysis}.
    % \ming{Need to revisit and update these numbers.}
        \label{tb:ccoeffs}}
    \begin{ruledtabular}
    \begin{tabular}{cccccc}
    &   &  Axial modes &   &  Horizontal modes  &  Vertical modes \\
    $\omega_p/2\pi$ & $150~$kHz & $600~$kHz & $2.0~$MHz & $3.0~$MHz & $2.5~$MHz \\
    \hline
    $|c_{p,y_b}^\beta|\times\sqrt{N}$ & $\lesssim 1$e-$2$ & $\lesssim 6$e-$3$ & $\lesssim 3$e-$3$ & $\lesssim 6$e-$2$ & $\lesssim 3$e-$3$ \\
    $|c_{p,x_b}^\gamma|\times\sqrt{N}$ & $\lesssim 1$e-$2$ & $\lesssim 7$e-$3$ & $\lesssim 4$e-$3$ & $\lesssim 2$e-$4$ & $\lesssim 2$e-$4$ \\
    $|c_{p,z_b}^\gamma|\times\sqrt{N}$ & $\lesssim 1$e-$4$ & $\lesssim 7$e-$5$ & $\lesssim 4$e-$5$ & $\lesssim 3$e-$5$ & $\lesssim 7$e-$4$ \\
    $|c_{p,x_b}^\lambda|\times\sqrt{N}$ & $\lesssim 7$e-$5$ & $\lesssim 4$e-$5$ & $\lesssim 2$e-$5$ & $\lesssim 3$e-$4$ & $\lesssim 2$e-$5$ \\
    $|c_{p,z_b}^\lambda|\times\sqrt{N}$ & $\lesssim 4$e-$6$ & $\lesssim 2$e-$6$ & $\lesssim 1$e-$6$ & $\lesssim 2$e-$5$ & $\lesssim 9$e-$7$ \\
    \end{tabular}
    \end{ruledtabular}
\end{table*}

\subsection{Hamiltonian approximation}
\label{sec:approx}

Equipped with the realistic parameter values detailed above, we now proceed with the power-series truncation strategy laid out in Sec.~\ref{sec:channel}. To do so, we need to first determine the extent of truncation in the motional Hilbert space. As a guiding principle, we would like to include a large enough
motional space for a specific mode so that for a thermal state considered
in that mode the population distributed outside of the
truncated motional space accounts for less than $10^{-3}$.
For the non-axial directions, we base this off of the initial temperature of our ion crystal, which is the Doppler temperature \footnote{For sideband cooled
horizontal modes in counter-propagating set up, see discussion
on Debye-Waller effect in Sec.~\ref{sec:approx}},
since the crystal does not easily heat in these directions.
For the axial, the modes readily heat, and we are interested in 
the fidelity impact from excessive axial mode temperature after 
a time period of heating.
%in how quickly they heat. \nam{@Ming: Insert the 1 and 10 quanta calculations for the non-axial directions after cooling here. Add one setence after that about the intent to capture the heating up to about a hundred quantas in the axial direction.} 
This motivates us to consider $\sim 10^2$ quanta for the non-axial and $\sim 10^4$ quanta for the axial cutoffs for the three directions, assuming e.g., each and every mode for a given direction heat more or less evenly. Note however that it is possible that there could be a dominant mode per direction that heats the most while the rest of the modes do not readily heat. To account for such a case, we also consider $\sim 10^2\times N$ quanta for the dominant non-axial and $\sim 10^4\times N$ quanta for the dominant axial modes. When determining which
operator terms $\hat{O}_{ij}$ to drop from our Hamiltonian, we consider both cases, i.e., even heating of all modes and concentrated heating of a dominant mode for each direction. We drop $\hat{O}_{ij}$ from the Hamiltonian in devising the effective Hamiltonian, only if the fractional contribution from $\hat{O}$ is less than $10^{-2}$ in both cases. 
We assume $N \leq 50$ for concreteness.

\comment{
\nam{--- old ---}
A first step in properly truncating the series in Eqs.~(\ref{eq:ABalternative}) is to evaluate the contribution from an operator $\hat{O}$, such as ..., \nam{@Ming: At this point, I think we need to specify what these operators are. Please insert the necessary detail here.} 
\ming{@Nam: If we say $H_I = (\sum_i \hat{O}_i)\hat{\sigma}+$h.c., and $\hat{O}_i$ 
can be written as const$\times \Pi \lambda_i^n \gamma_j^m \beta_k^l$ then
$\hat{O}$ is each one of the $\hat{O}_i$. I don't know how to 
express this easily.}
in the Hamiltonian by the overlap metric ${\mathcal E}(\hat{O})$ defined in (\ref{eq:errmetric}). 
In order to evaluate the metric ${\mathcal E}(\hat{O})$ defined in (\ref{eq:errmetric}), we need to choose $n_1^{\mathrm{max}},\cdots ,n_{3N}^{\mathrm{max}}$. 
We chose ... \nam{@Ming: Insert what values were actually used for our calculations} 
\ming{@Nam: So these depends on which case we consider, which are given later}
for the following reasons.
\nam{@Ming: Is there a reason why we provide two trivial bounds?}
\ming{@Nam: not really. Just that these are the cases I used to evaluate the terms}
For a given total energy in the motional space, consider 
the two ends of the spectrum of energy partitions among motional modes. As will be clear, all other cases lay in between the two. 
One end of the spectrum we consider is the case where only one mode in a set of $N$ modes along the same direction has non-zero occupation. 
We assume based on ... \nam{@Ming: Where did these formulas come from?} 
\ming{@Nam: These are basically $10\times$ the maximal phonon number we
see for each set of mode then times $N$ because the heating naively scales
linearly with $N$.}
that %for the dominant horizontal mode
$n^{\mathrm{max}}_{\mathrm{H}}=10\times N$,% for the dominant vertical mode
$n^{\mathrm{max}}_{\mathrm{V}}=100\times N$, and% for the dominant axial mode
$n^{\mathrm{max}}_{\mathrm{A}}=2000\times N$, for the dominant modes in each of the horizontal, vertical, and axial directions, respectively.
\ming{Double check these numbers later after the experimental analysis.
Also, this is just so that we can check experimental results given later.
In realistic situation, the temperature should be quite low with small
error rate. Do we want to explain this point?}
On the other end of the spectrum, we assume that all modes along a given direction contribute equally and thus they have the same occupation. In this case, we use for each mode in the respective directions
%the average occupation for the horizontal modes
$n^{\mathrm{max}}_{\mathrm{H}}=10$, %for the vertical modes to be
$n^{\mathrm{max}}_{\mathrm{V}}=100$, %and for the axial modes to be
$n^{\mathrm{max}}_{\mathrm{A}}=2000$. 
\nam{@Ming: What purpose does the statement $N \le 50$ serve here?}
\ming{@Nam: For the case where heating is evenly distributed, N is not trivially
canceled out by the $\nu$ terms because the first order contribution is
proportional to $1/\sqrt{N}$ times $N$ mode. Then $N$ becomes important
in evaluating the fractional contribution of each term. We can chat about
this via a call.}
For all situations, we assume $N \le 50$.
}

The expressions in (\ref{eq:ABfunc}) (see Appendix~\ref{sec:app_p1}, Eq.~(\ref{eq:ABalternative}) for the ordered form)
%\nam{Do we have to start from these or can we just say we start from the A/B coefficients given in Sec. III, Eq. (15)?}
%\ming{We can say from Eq. (15), and more specifically the ordered form from A1.}
%\nam{I suggest we use (15), not (A1)}
may now be approximated according to the strategy outlined in Sec.~\ref{sec:channel} with the parameters specified in Sec.~\ref{sec:spec}. Keeping only the terms with the size of the fractional contribution larger than $10^{-2}$, we obtain 
%\nam{@Ming, please use $\lambda$ and $\gamma$ throughout from here on, instead of $p$ and $q$}
%\ming{@Nam: $p$ and $q$ can be $\lambda$ and $\gamma$ with $x_b$ or $z_b$
%subscripts. Maybe keep the p and q for Eq 17?}
\begin{align}
    A_1 \approx& \; s_0^{1/2}\;, \nonumber \\
%%%%%%%%%%%%%%%%%%%%%%%%%%%%%%%%%%%%%%%
    A_2 \approx& \;
      e^{-s_0^2 q_0^2}\sum_{l_q=0}^{\infty}
      \frac{(-s_0 \hat{q}_1)^{l_q}}{{l_q!}}
        \mathcal{H}_{l_q}(s_0 q_0) \nonumber \\
        =& \;
      \exp\left[-s_0^2(q_0 + \hat{q}_1)^2\right] \nonumber \\
    %   \approx & \;
    %   \exp(-{\gamma_{z_b}^{(0)}}^2/(1+{\lambda_{z_b}^{(0)}}^2))\;, \nonumber \\
%%%%%%%%%%%%%%%%%%%%%%%%%%%%%%%%%%%%%%%
    B_0^{\pm} =& \; e^{\mp i \hat{p}_1}, \nonumber \\
%%%%%%%%%%%%%%%%%%%%%%%%%%%%%%%%%%%%%%%
    B_1^{\pm} \approx& \; e^{\pm \frac{i}{2} \arctan (p_0)} \;, \nonumber \\
%%%%%%%%%%%%%%%%%%%%%%%%%%%%%%%%%%%%%%%
    B_2^{\pm} \approx& \; e^{\mp i s_0^2 q_0^2 p_0} \;,
\label{eq:ABsimplified}
\end{align}
where $s_0 = 1/\sqrt{1+p_0^2}$.
Note we may further approximate the 
the $A_2$ function in (\ref{eq:ABsimplified}) 
in the case where it is used for $z_b$ direction,
as in the second $A_2$ function used in (\ref{eq:Eb}).
Specifically, $A_2(\lambda_{z_b}^{(0)},\hat{\lambda}_{z_b},
        \gamma_{z_b}^{(0)},\hat{\gamma}_{z_b})$ function
may be truncated to 
$\exp(-{\gamma_{z_b}^{(0)}}^2/(1+{\lambda_{z_b}^{(0)}}^2))$, 
due to the larger beam waist and lower vertical mode temperature
along the $z_b$ directions.
We assumed here 
$q_0 = \gamma_{z_b}^{(0)}$ and 
$\hat{q}_1 = \hat{\gamma}_{z_b}$.
%\nam{@Ming: the beam waist etc. consideration are all a part of the process in keeping %only the terms with the size less than $10^{-2}$ correct?}
%\ming{Yes, they are all used to truncate to $10^{-2}$ accuracy.}

%\nam{@Ming, is this what we use for $A_2$ or are we stating this for some other purposes? Also, why is it that we can drop $\hat{q}_1$ if we assume $q_0 = \gamma_{z_b}^{(0)}$ and $\hat{q}_1 = \hat{\gamma}_{z_b}$?}
%\ming{@Nam, this is only the case for $z_b$ axis (vertical), not for $x_b$ axis,
%because of its larger waist and lower temperature. I've done
%some modification there, please take a look.}
%Before we finally present the simplified interaction Hamiltonian,
%there is one more approximation we can do.
%\nam{@Ming, the $B_0$ approximation: This is actually not the approximation of $B_0$ itself per se -- rather it sounds like you're approximating here, knowing that things will cancel out when you consider the two beams for co- or counter-propagating beams, correct?}
%\ming{I restored equation 17. Do you mind rework the wordings in the following
%paragraph please?}
%\nam{--- to here ---}

Inserting the simplified $A$ and $B$ functions
to the amplitude and phase functions in (\ref{eq:Eb}) and (\ref{eq:Phib}), respectively,
then inserting the simplified amplitude and phase functions to the interaction Hamiltonian in (\ref{eq:HI}), 
we obtain
\begin{widetext}
\begin{equation}
    % H_I = \hbar\Omega_0 \sum_{n=0}^{\infty}\sum_{m=0}^{n} &\;
    %  \frac{(-1)^n}{m!(n-m)!}
    %  \left(\frac{\hat{\gamma}_{x_1}}{\sqrt{1+{\lambda_{x_1}^{(0)}}^2}}\right)^m
    %  \left(\frac{\hat{\gamma}_{x_2}}{\sqrt{1+{\lambda_{x_2}^{(0)}}^2}}\right)^{n-m} \nonumber \\
    %  & \mathcal{H}_m\left(\frac{\gamma_{x_1}^{(0)}}{\sqrt{1+{\lambda_{x_1}^{(0)}}^2}}\right)
    %  \mathcal{H}_{n-m}\left(\frac{\gamma_{x_2}^{(0)}}{\sqrt{1+{\lambda_{x_2}^{(0)}}^2}}\right)
    %  \left[e^{i\Delta\omega t + \Psi_0}e^{i(\hat{\beta}_2-\hat{\beta}_1)}
    %  \hat{\sigma}^+ + \mathrm{h.c.}\right] \;,
    H_I = \hbar\Omega_0 \sum_{l=0}^{\infty}\sum_{m=0}^{l}
     \frac{(-1)^l}{m!(l-m)!}
     {\hat{\gamma}_{\lambda,x_1}}^m
     {\hat{\gamma}_{\lambda,x_2}}^{l-m}
     \mathcal{H}_m\left(\gamma_{\lambda,x_1}^{(0)}\right)
     \mathcal{H}_{l-m}\left(\gamma_{\lambda,x_2}^{(0)}\right)
     \left[e^{i[(\omega^{\mathrm{qbt}}+\Delta\omega)t + \Psi_0]}
     \hat{\sigma} + \mathrm{h.c.}\right] \;,
\label{eq:HIaxial}
\end{equation}
\end{widetext}
with $\hat{\gamma}_{\lambda,x_b}\hspace{-0.4em}=
\hspace{-0.2em}\hat{\gamma}_{x_b}/\sqrt{1+{\lambda_{x_b}^{(0)}}^2}$
and $\gamma_{\lambda,x_b}^{(0)}\hspace{-0.4em}=
\hspace{-0.2em}\gamma_{x_b}^{(0)}/\sqrt{1+{\lambda_{x_b}^{(0)}}^2}$.
We collected all non-operator terms into a Rabi rate term $\Omega_0$
and a phase term $\Psi_0$, defined as
% \vspace{10pt}
% \begin{widetext}
\begin{align}
    \Omega_0 =&\; \frac{\bar{D}}{\pi\hbar} \sqrt{\frac{P_1P_2}
     {w_{x_1}^{\mathrm{f}}w_{x_2}^{\mathrm{f}}w_{z_1}^{\mathrm{f}}w_{z_2}^{\mathrm{f}}}} \nonumber\\
     &\hspace{-2.25em}\left[(1+{\lambda_{x_1}^{(0)}}^2)(1+{\lambda_{x_2}^{(0)}}^2)
     (1+{\lambda_{z_1}^{(0)}}^2)(1+{\lambda_{z_2}^{(0)}}^2)\right]^{-1/4} \nonumber \\
     &\hspace{-2.25em}\exp\hspace{-0.2em}
     \left[-\frac{{\gamma_{x_1}^{(0)}}^2}{1+{\lambda_{x_1}^{(0)}}^2}
     -\frac{{\gamma_{x_2}^{(0)}}^2}{1+{\lambda_{x_2}^{(0)}}^2}
     -\frac{{\gamma_{z_1}^{(0)}}^2}{1+{\lambda_{z_1}^{(0)}}^2}
     -\frac{{\gamma_{z_2}^{(0)}}^2}{1+{\lambda_{z_2}^{(0)}}^2}
     \right]\;
\end{align}
and
\begin{align}
    \Psi_0 =&\; \phi_1 - \phi_2 + k_2 y_{x_2}^{(0)} - k_1 y_{x_1}^{(0)} \nonumber \\
    &\hspace{-2.25em} +\frac{1}{2}\left[\arctan\lambda_{x_1}^{(0)}\hspace{-0.2em}+
     \arctan\lambda_{x_2}^{(0)}\hspace{-0.2em}+
     \arctan\lambda_{z_1}^{(0)}\hspace{-0.2em}+
     \arctan\lambda_{z_2}^{(0)}\right] \nonumber \\
    &\hspace{-2.25em} 
    -\left[\frac{\lambda_{x_1}^{(0)}{\gamma_{x_1}^{(0)}}^2}{1+{\lambda_{x_1}^{(0)}}^2}
    +\frac{\lambda_{x_2}^{(0)}{\gamma_{x_2}^{(0)}}^2}{1+{\lambda_{x_2}^{(0)}}^2}
    +\frac{\lambda_{z_1}^{(0)}{\gamma_{z_1}^{(0)}}^2}{1+{\lambda_{z_1}^{(0)}}^2}
    +\frac{\lambda_{z_2}^{(0)}{\gamma_{z_2}^{(0)}}^2}{1+{\lambda_{z_2}^{(0)}}^2}
    \right] \;.
\end{align}
% \end{widetext}
To arrive at (\ref{eq:HIaxial}), we used
$\|e^{i(\hat{\beta}_2-\hat{\beta}_1)}-1\| \approx 1$
%to within $10^{-2}$ 
for both the co- and counter-propagating beams, which
means that the Debye-Waller effect~\cite{NISTBible} is negligible.
In the co-propagating set up, the local coordinate systems of the beams are
mostly aligned and when the two $B_0^\pm$ functions are substituted into
the Hamiltonian, the resulting term $e^{i(\hat{\beta}_2-\hat{\beta}_1)}$
becomes very close to an identity operation due to the cancellation of the 
$\hat{\beta}_b$ operators. More specifically, using the information of the 
system given previously, it is straightforward to show that 
$\|e^{i(\hat{\beta}_2-\hat{\beta}_1)}-1\|$ is smaller than $10^{-2}$.
In the counter-propagating set up, $e^{i(\hat{\beta}_2-\hat{\beta}_1)}-1$
does contribute significantly to the Hamiltonian, 
% which is known as the Debye-Waller effect~\cite{NISTBible}, 
unless the horizontal modes are sufficiently cooled so that $n_H \ll 1$.
%For specific experiments where we use the counter-propagating set up,
Since we cool the horizontal modes using sideband cooling to suppress
the Debye-Waller effect in all experiments that use the counter-propagating
set up, $\|e^{i(\hat{\beta}_2-\hat{\beta}_1)}-1\|$ may be assumed 
smaller than $10^{-2}$ and can thus be neglected, 
like in the co-propagating case.

The interaction Hamiltonian in (\ref{eq:HIaxial}) can readily be used to
assess fidelity impacts of noise sources ranging from beam misalignment 
and instability, to noise on the ion positions, as well as excessive mode 
temperature in a single-qubit gate operation using, for instance, a 
Monte-Carlo type simulation. It can also be easily incorporated into a two-qubit
Hamiltonian to evaluate errors in a two-qubit gate operation.
Note that when the two beams are perfectly aligned with each other, Eq.~(\ref{eq:HIaxial})
reduces to a single summation with only one Hermite polynomial term in each
summand, which reproduces the results given in Ref.~\cite{WesAxial}.

\subsection{Axial mode temperature effect}
\label{sec:Axial}

%\ming{@Nam: could you double check the following paragraph?}
Here, we put our Hamiltonian expression in (\ref{eq:HIaxial})
to test by investigating the fidelity impact of excessive 
axial mode temperature. 
% \ming{Kristi: Suggestion: I'd add another flagging phrase or sentence here to remind the reader that this is just one application of the Hamiltonian derived in III before proceeding to the assumptions, or add a paragraph break after the first sentence.}
We assume that only one axial mode, for instance
the center-of-mass (COM) mode, has a dominant behavior in determining
the temperature,
%and neglect all other mode ladder operators, 
thus dropping the mode index $p$.
Next, we consider two realistic situations regarding the beam waist and
the beam alignment. 

In the first situation, which is representative of
the co-propagating set up, we have
two tightly focused beams with identical waists 
$w_{x_1}^\mathrm{f} = w_{x_2}^\mathrm{f}$. As a good approximation,
we can assume that they are perfectly aligned, {\it i.e.}
$\gamma_{\lambda,x_1}^{(0)} = \gamma_{\lambda,x_2}^{(0)}$
and $|c_{p,x_1}^{\gamma}| = |c_{p,x_2}^{\gamma}|$.
Then the summation of $m$ in (\ref{eq:HIaxial}) reduces to a single term
due to a sum rule of products of two Hermite polynomials.
% perfect alignment between the two beams and 
% drop the subscript of the beam index $b$.
% \ming{Still working here}
We can then transform the interaction 
Hamiltonian with $U_0=\exp{(-iH_0t/\hbar)}$ and obtain
\begin{widetext}
\begin{equation}
    H_I^\prime = \hbar\Omega_0 \sum_{l=0}^{\infty}\frac{(-1)^l}{l!}
     \left(\eta\right)^l  \mathcal{H}_l\left(\xi\right)
     (e^{-i\omega t}\hat{a}+
     e^{i\omega t}\hat{a}^\dag)^l
     \left[e^{i(\Delta\omega t + \Psi_0)}
     \hat{\sigma} + \mathrm{h.c.}\right] \;,
    \label{eq:HIprime}
\end{equation}
\end{widetext}
where
\begin{align}
    \eta =&\; \frac{\zeta^{(0)}\nu^{x}}
      {w_{x}^{\mathrm{eff}}\sqrt{1+{y_x^{(0)\mathrm{f}}}^2/{y^{\mathrm{R}}_x}^2}} \;,
      \nonumber \\
    \xi =&\; \frac{x^{(0)}}
      {w_{x}^{\mathrm{eff}}\sqrt{1+{y_x^{(0)\mathrm{f}}}^2/{y^{\mathrm{R}}_x}^2}} \;.
\label{eq:etaxi}
\end{align}
$w_{x}^{\mathrm{eff}}$ is the effective waist given by
$w_{x}^{\mathrm{f}}/\sqrt{2}$. 

In the second situation, which is representative of
the counter-propagating set up, we have one of the Raman beams to be
narrowly focused and individually addressing while the other
to be very loosely focused and capable of addressing a long ion chain.
The loosely focused, global addressing beam has a waist of more than
$100~\mu$m which allows us to truncate any $\hat{\gamma}_{\lambda,x}^m$
term with $m>0$. Thus, the interaction Hamiltonian is
again of the form in (\ref{eq:HIprime}). The only difference is that
the effective waist here is given by waist $w_{x}^{\mathrm{f}}$ of the narrowly
focused beam.

For a single-qubit gate operation, we drive the Raman transition at the qubit
frequency, {\it i.e.} $\Delta\omega = 0$. Then any term with imbalanced
numbers of $\hat{a}$ and $\hat{a}^\dag$ are off-resonant and thus suppressed
\footnote{Note that the terms with odd $l$ in Eq.~(\ref{eq:HIprime}) are
suppressed by the alignment parameter $\xi$ due to the fact that the
zeroth order terms of the odd Hermite polynomials are on the order of
$\xi$ instead of a constant. Therefore, the coefficients
in front of the imbalanced terms are one $\xi$ order smaller than
the balanced terms.}. 
Neglecting these fast-rotating couplings, we can simply write the evolution
operator of a single-qubit gate pulse that has a constant power of a time 
duration of $t_{\mathrm{sqg}}$ as
\begin{equation}
    U_I = e^{-i H_I^\prime t_{\mathrm{sqg}}/\hbar} \approx 
    \sum_{n} \left(\cos\Theta_{n} \hat{I}
    - i\sin\Theta_{n} \hat{\sigma}_{\Psi_0}\right)|n\rangle\langle n| \;,
\label{eq:sqgunitary}
\end{equation}
where $|n\rangle$ is a Fock state in the axial mode space, 
%$\hat{\sigma}_{\Psi_0} = \cos(\Psi_0)\hat{\sigma}^{x} + \sin(\Psi_0)\hat{\sigma}^{y}$,
$\hat{\sigma}_{\Psi_0} = \exp(i\Psi_0)\hat{\sigma} + \mathrm{h.c.}$,
$\hat{I}$ is the identity operator in the qubit space, and $\Theta_n$
is defined as
\begin{equation}
%  \Theta_n = \Omega_0 t\;\langle n | \sum_{l=0}^{\infty} \frac{(\eta)^{2l}}{(2l)!}
%     \mathcal{H}_{2l}(\xi) (\hat{a}+\hat{a}^\dag)^{2l} | n \rangle \;.
 \Theta_n = \Omega_0 t_{\mathrm{sqg}}\; \sum_{m=0}^{\infty}
  \left(\frac{-\eta^2}{2}\right)^{m}\frac{\mathcal{H}_{2m}(\xi)}{m!}
    \vphantom{1}_2\mathcal{F}_1(1+n,-m;1;2)  \;.
\label{eq:Theta}
\end{equation}
Here we used
\begin{align}
\hspace{-0.5em}
   \langle n | (\hat{a}+\hat{a}^\dag)^{2m} | n \rangle =&\, \sum_{i=0}^m
     \left(-\frac{1}{2}\right)^{\hspace{-0.2em}m-i}\hspace{-0.2em}
     \frac{(n+i)!\,(2m)!}{n!\,(m-i)!\,(i!)^2} \nonumber \\
     =&\, \left(-\frac{1}{2}\right)^{\hspace{-0.2em}m}
     \frac{(2m)!}{m!}
     \vphantom{1}_2\mathcal{F}_1(1+n,-m;1;2) \;,
\label{eq:thermalExp}
\end{align}
%\ming{@Nam, could you double check the above equation?}
%\nam{Checked.}
where $\vphantom{1}_2\mathcal{F}_1(a,b;c;z)$ denotes a 
Gaussian hypergeometric function.
Equation (\ref{eq:Theta}) explicitly shows how the Rabi
frequency for driving the spin degree of freedom depends on
the phonon occupation of the axial motional state.
Thus a distribution of the phonon occupation number of the
axial motional state with a non-zero width results in 
a distribution of the Rabi frequency with a corresponding
non-zero width which in turn induces decoherence
to the quantum gate operation.

Note that the convergence of (\ref{eq:Theta}) greatly depends on
$\eta$ and $n$. For instance, for perfect alignment, {\it i.e.}
$\xi \to 0$, with $\eta = 0.01$ and
$n = 2000$, we need $m=4$ to achieve convergence to the
third significant digit. To achieve the same accuracy, we need $m=11$
for $\eta = 0.02$ and $n = 2000$, and $m=92$ for
$\eta = 0.02$ and $n = 20000$.
% \ming{I think a convergence graph with respect to $m$ for specific 
% values of $\xi$, $n$, and $\eta$ would be interesting. Just to show
% how many orders do people need to take (in comparison with Marco's paper
% which is explicitly $m=1$). The counter point is that higher phonon 
% number is not what an error analysis focuses on.
% So maybe there's no point showing that. What do you guys think?}
To mitigate some of the convergence issue,
if we assume  $\xi \to 0$,
we have $\mathcal{H}_{2m} = (-1)^m(2m)!/m!$, 
and we can simplify (\ref{eq:Theta}) to
\begin{equation}
  \Theta_n = \frac{\Omega_0 t_{\mathrm{sqg}}}
    {\sqrt{1+2\eta^2}} \,
    \vphantom{1}_2\mathcal{F}_1(\frac{1}{2},-n;1;\frac{4\eta^2}{1+2\eta^2}) \;.
\label{eq:ThetaSimp}
\end{equation}
Once proper care for convergence is taken,
it is straightforward to insert (\ref{eq:ThetaSimp}) in (\ref{eq:sqgunitary})
to evaluate the effect of the axial mode temperature on the
fidelity of a single-qubit gate operation for different
initial states and measurement schemes. We carry
out such an analysis in more detail in the next section 
in conjunction with our experimental results.

%\subsection{Co-propagating Raman beam set-up}

\section{Experiment}
\label{sec:experiment}

In this section, we experimentally investigate the impact of high-temperature axial modes in the presence of tightly focused Raman laser beams.
A similar experimental apparatus has been discussed in detail elsewhere~\cite{benchmarkPaper}, but we briefly discuss the key features here for completeness.
We load a chain of $^{171}$Yb$^+$ ions in a surface-electrode ion trap, where we can control the axial chain spacing by adjusting the voltages on several DC electrodes on the trap.
% Quantum logic gates are performed via Raman laser interaction, where we achieve individual qubit addressability using a combination of a global addressing Raman laser beam and an array of tightly focused laser beams derived from a common pulsed 355-nm laser.
% The individual Raman beams have a $1/e^2$ waist of $0.99(2)$~um along the chain axis, while the global beam has a $1/e^2$ waist of $\approx 200$~um.
Quantum logic gates are performed via Raman transitions induced by
two $355~$nm Gaussian beams.
The state initialization follows a Doppler cooling sequence, where
the initial motional temperature is cooled to the Doppler limit.
When using the counter-propagating set up, the horizontal modes are
further cooled to $\bar{n}\lesssim 0.1$ using a sideband cooling sequence.
This reduces to the effective Hamiltonian of Eq.~(\ref{eq:effectiveH0}), where our qubit states $\{|\hspace{-0.25em}\downarrow\rangle, |\hspace{-0.25em}\uparrow\rangle\}$ are taken to be the $|F=0, m_F=0\rangle$ and $|F=1,m_F=0\rangle$ hyperfine levels of the ground electronic state, respectively.
High-fidelity state preparation is done via optical pumping to the $|\hspace{-0.25em}\downarrow\rangle$ at the beginning of each experiment, and measurement is done by spatially resolved, state-dependent fluorescence detection~\cite{SOlmschenk2007}.

% \begin{itemize}
%     \item Evidence of fidelity decay due to axial heating
%     \item Optimized Rabi scheme increase fidelity
%     \item Using the fidelity decay as a way to measure heating rate
%     \item Effect of pulse sequences
% \end{itemize}

\subsection{Measurement of axial mode temperature effect}

\begin{figure}
    \centering
    \includegraphics[width=1.05\columnwidth]{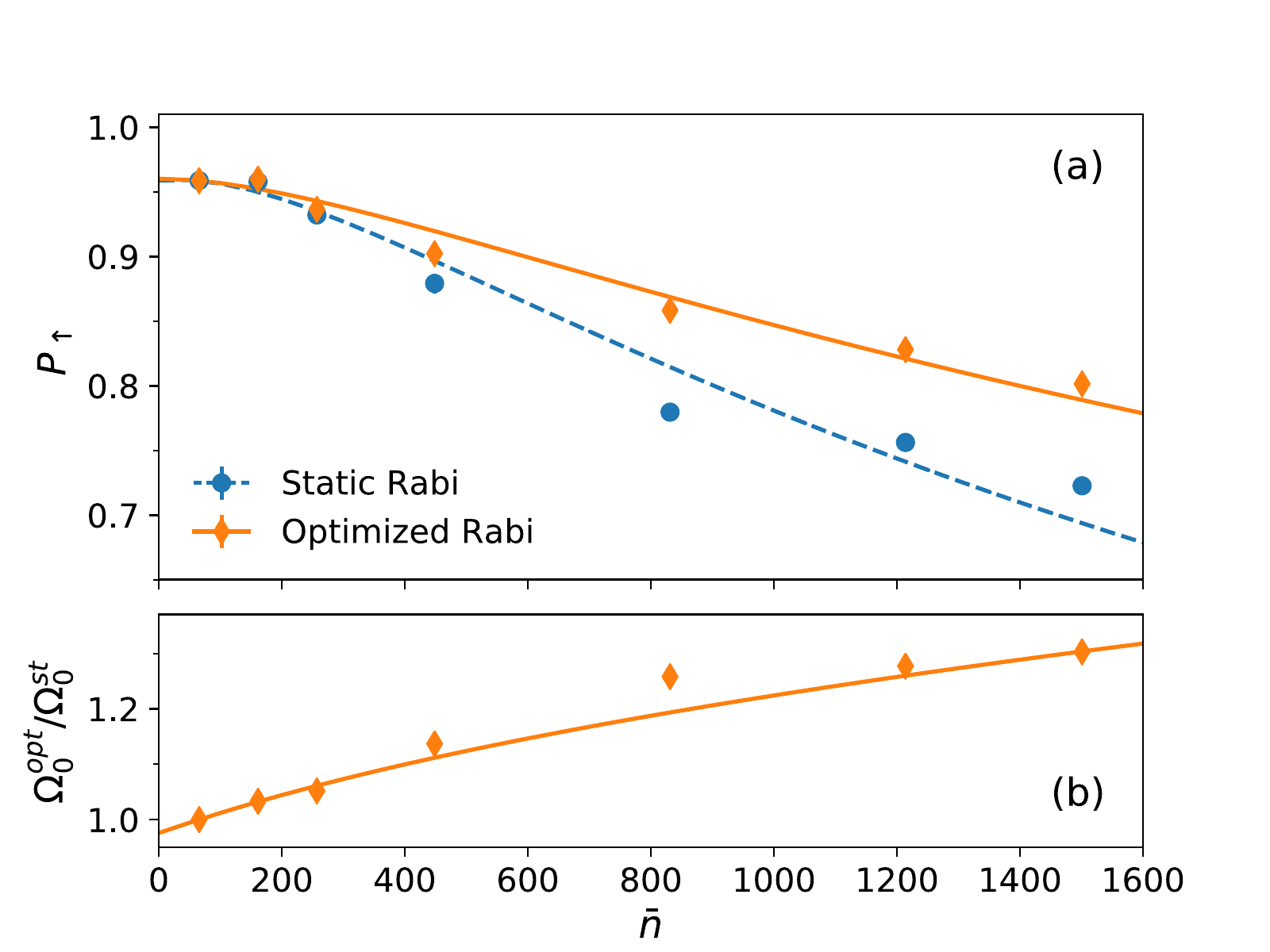}
    \caption{Implementation of the delayed gate protocol (A)-(D) detailed in the main text.
    A single ion with axial frequency $\omega_A = 2\pi\times153~$kHz
    was used in a co-propagating set up. (a) $P_{\uparrow}$ as a function
    of $\bar{n}$. Experimental (points) and simulation 
    (lines) results show the optimized Rabi approach (orange diamonds and solid line) can
    mitigate the $P_{\uparrow}$ decay induced by the mode heating,
    shown here by the static approach result (blue circles and dashed line).
    (b) the ratio $\Omega_0^{\rm opt}/\Omega_0^{\rm st}$ as a function of $\bar{n}$.
    The orange diamonds are the experimental 
    results and the orange solid line is the theoretical simulation 
    results. 
    See Sec.~\ref{sec:probe} for model details.
    % \ming{Kristi: Unless I missed it, the low $P_{\uparrow}(0)$ is never addressed directly for this figure. I'd mention it either in the first discussion of the figure (bottom of LHS column on page 9) or as an explanation for the added $\delta P_{\uparrow}$ term in equation 29. }
    }
    \label{fig:1ion153}
\end{figure}

\begin{figure}
    \centering
    \includegraphics[width=1.05\columnwidth]{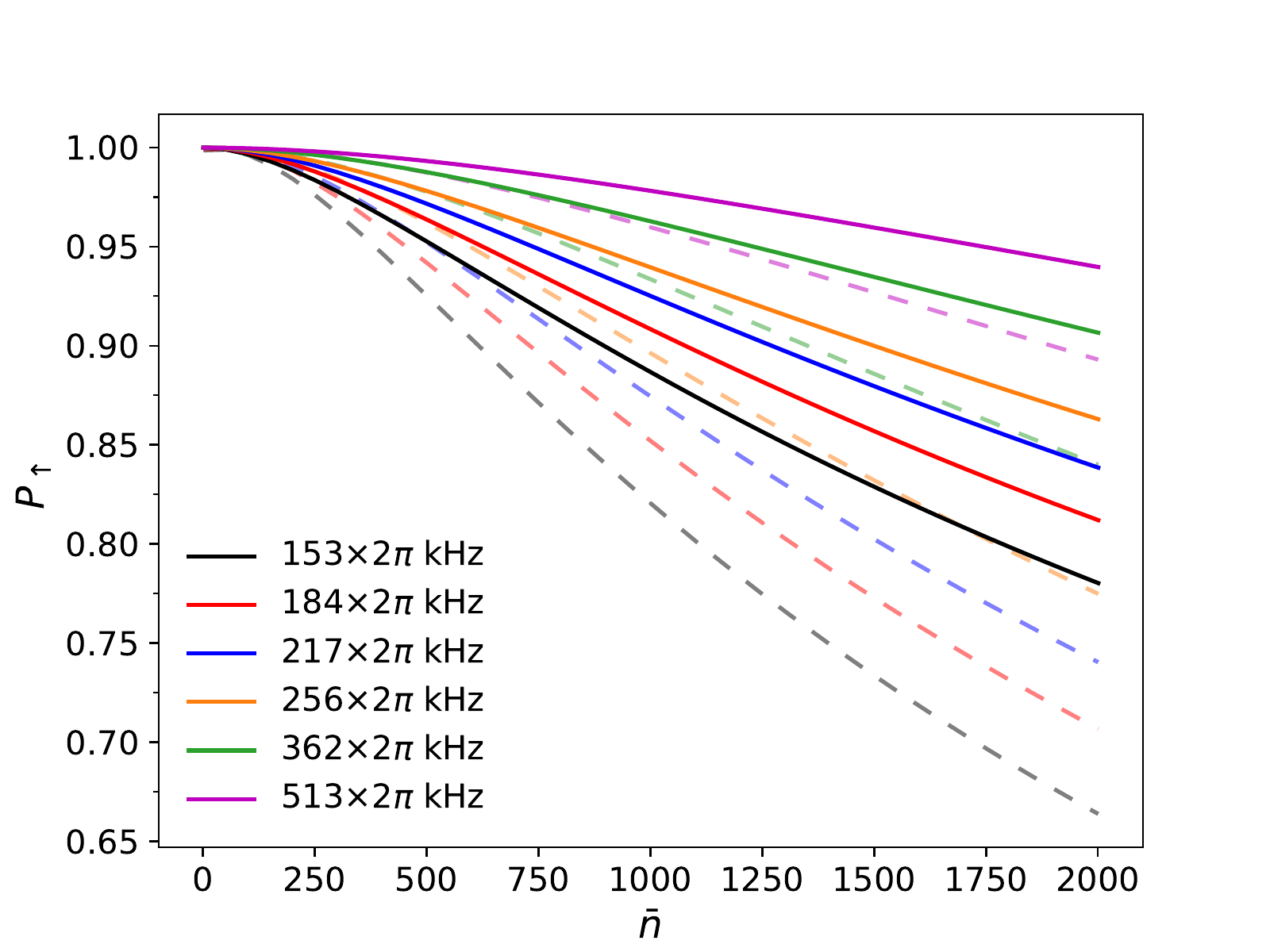}
    \caption{Theoretical simulations of $P_{\uparrow}$ as a function
    of $\bar{n}$ for single ion in a co-propagating set up. 
    Results for six different axial mode frequencies are presented.
    The solid lines represent the optimized Rabi approach and the
    dashed lines represent the static Rabi approach.\
    For each line style, the color coded lines from top to bottom 
    correspond to having axial mode frequencies from high to low as
    listed in the legend.}
    \label{fig:1ionsim}
\end{figure}

Following the theoretical analysis shown in Sec.~\ref{sec:error_analysis}, we experimentally probe the following steps specifically:
(A) we initialize our quantum state to
$\rho_0 (0)=|\hspace{-0.25em}\downarrow\rangle\langle\downarrow\hspace{-0.25em}|
\otimes\rho_T(0)$,
where $|\hspace{-0.25em}\downarrow\rangle$ is a qubit state vector and 
\begin{equation}
  \rho_T(t)=
  \sum_n \frac{\bar{n}_t^n}{(1+\bar{n}_t)^{n+1}} |n\rangle\langle n|
\end{equation}
is the density operator of a thermal state of the axial mode at time $t$
with an average Fock state occupation number $\bar{n}_t$;
(B) we wait for time $\Delta t$ so that the axial motional state
is heated to a higher $\bar{n}_{\Delta t}$ and the quantum state
becomes $\rho_0 (\Delta t)$;
(C) we implement a single-qubit gate whose unitary is given by 
(\ref{eq:sqgunitary}) with $\hat{\sigma}_{\Psi_0} = \hat{\sigma}_x$;
(D) we measure the final state and
repeat the experiment to sample the probability of
the final state being measured in $|\hspace{-0.25em}\uparrow\rangle$. 
Denoting the measurement projector as 
$M = |\hspace{-0.25em}\uparrow\rangle\langle\uparrow\hspace{-0.25em}|
\otimes \hat{I}_{\mathrm{mot}}$, where $\hat{I}_{\mathrm{mot}}$ is the identity
operator in the motional space, based on Sec.~\ref{sec:Axial}, the probability to measure a positive measurement outcome $P_{\uparrow}$ is given by
\begin{align}
    P_{\uparrow}(\bar{n}_{\Delta t}) &= \mathrm{Tr}[U_I \rho_0(\Delta t) U_I^\dag M]
      \nonumber \\
      &= \sum_n \frac{\bar{n}_{\Delta t}^n}{(1+\bar{n}_{\Delta t})^{n+1}}
      (\sin\Theta_n)^2 \;,
\label{eq:pbright}
\end{align}
% \nam{Need to explicitly state that this is the proxy we use to represent single-qubit gate fidelity.}
where $\bar{n}_{\Delta t}$ is the average phonon number after heating over
the duration $\Delta t$. We use delay times on the order of ms which is
much larger than the duration of single-qubit gate operations
which are on the order of $10$ to $100~\mu$s. Thus we neglect the
heating during the gate operation.
Here the bright population $P_{\uparrow}$ is a direct measure 
of the final state fidelity and hence a good proxy metric for the single
qubit gate fidelity.
We repeat the same set of steps for multiple
$\Delta t$ values for each experimental set up of different
beam arrangements, chain lengths, as well as axial frequencies.

Figure~\ref{fig:1ion153} shows the bright population $P_{\uparrow}$ 
as a function of the average number of phonons $\bar{n}$, 
obtained according to the expression (\ref{eq:pbright}) 
for the axial frequency $2\pi\times153{\rm kHz}$.
Specifically, we optimize $P_{\uparrow}$ in (\ref{eq:pbright}) 
with respect to $\Omega_0$, while assuming the initial average phonon number 
is $\bar{n}_0 \approx 64$, whose value is commensurate
to the Doppler limit of the axial mode. 
Once the specific $\Omega_0$ is obtained, hereafter referred to
as the static Rabi rate $\Omega_0^{\rm st}$, we may plot
$P_{\uparrow}$ as a function of $\bar{n}$.
To compare, experimentally, we measure $P_{\uparrow}$
as a function of the delay time $\Delta t$ using the
static Rabi rate $\Omega_0^{\rm st}$, calibrated without any delay,
and we map $\Delta t$ to $\bar{n}$ according to a constant heating rate
model, i.e., $\bar{n} = \bar{n}_0 + \dot{\bar{n}} \Delta t$.
The experiments were conducted on a single ion confined in a harmonic well, 
where the axial frequency was adjusted to $2\pi\times153$ kHz by changing 
the voltages of the DC electrodes of the ion trap.
The co-propagating beam set up was used with beam waists 
$w^{\mathrm{f}}_{x} = 1.4~\mu$m.
In Fig.~\ref{fig:1ion153}, we use $\dot{\bar{n}} \approx 96/{\rm ms}$, which
agrees with our model the best.
The agreement in $P_{\uparrow}$ decay 
between our static Rabi rate based model
and experimental results confirms the effect of heating
in the axial mode on the quantum gate fidelity.

\subsection{Improvement in the quantum gate fidelity}
\label{sec:Improve}

% \ming{Do we need to mention sympathetic cooling here?}
% \nam{No. We did not experimentally implement sympathetic cooling.}

Improvement in the the bright population 
$P_{\uparrow}$ hence the quantum gate fidelity over the static approach
may readily be achieved by the following.
Recall in our static approach we assumed the Rabi rate
to be that obtained for the average initial phonon number $\bar{n}_0$.
In theory, the bright population $P_{\uparrow}$ may be
maximized with respect to $\Omega_0$ for any $\bar{n}$.
If we thus allow for $P_{\uparrow}$ to be individually optimized for
different values of $\bar{n}$, we can obtain $P_{\uparrow}$ values
that are larger than those obtained by the static approach.
Figure~\ref{fig:1ion153} shows the optimized $P_{\uparrow}$ as a function 
of $\bar{n}$, which may be compared with the static counterpart. 
The experimental results are accordingly obtained by optimizing over 
the Rabi rate for each delay time $\Delta t$, mapped to $\bar{n}$ 
as described previously with the same heating rate $\dot{\bar{n}}$ and 
initial average phonon number $\bar{n}_0$.
By adjusting the Rabi rate according to $\bar{n}$, we achieve
improvement in the quantum gate fidelity.
The optimal Rabi rate obtained from this approach is hereafter
denoted as $\Omega_0^{\rm opt}$. 
Figure~\ref{fig:1ion153} (b) shows the agreement of the
ratio $\Omega_0^{\rm opt}/\Omega_0^{\rm st}$ between experimental
and simulated results. It is then possible, given a known
initial temperature and heating rate, to predict the optimal
Rabi rate for any quantum gate operation embedded in a quantum
circuit without explicit calibration, thus improve the
overall fidelity of the quantum circuit.

Further improvement in the quantum gate fidelity over
the axial mode heating may be achieved by raising the 
axial mode frequency through the following two mechanisms. 
Firstly, increasing the mode frequency $\omega_A$
decreases $\eta$ in (\ref{eq:etaxi}), which in turn
reduces the widths of the distribution of $\Theta_n$
with respect to a specific distribution of $n$ and
lessens its decoherence effect on the quantum gate operation.
Figure~\ref{fig:1ionsim} shows $P_{\uparrow}$ as a function 
of $\bar{n}$ for a variety of axial mode frequencies, ranging from
$2\pi\times153$ kHz to $2\pi\times513$ kHz. We observe that
both the static and optimized $P_{\uparrow}$ values
decay slower in $\bar{n}$ for higher frequencies.
For a given $\bar{n}$, a factor $R$ increase in the
axial mode frequency approximately translates to 
the gate infidelity reduction by $R$, which is due to the
fact that $1-P_{\uparrow}$ is proportional to $\eta^2$
to the zeroth order. Secondly, increasing an axial mode frequency
in most cases decreases the heating rate associated with
the mode~\cite{IBoldin2018} thus improves the overall 
fidelity of any quantum circuit of depth larger than one.

% The axial mode frequency may straightforwardly be adjusted
% by changing the electric potential of the electrodes
% of the ion trap in practice.

\subsection{Heating rate probe}
\label{sec:probe}

\begin{figure}
    \centering
    \includegraphics[width=1.05\columnwidth]{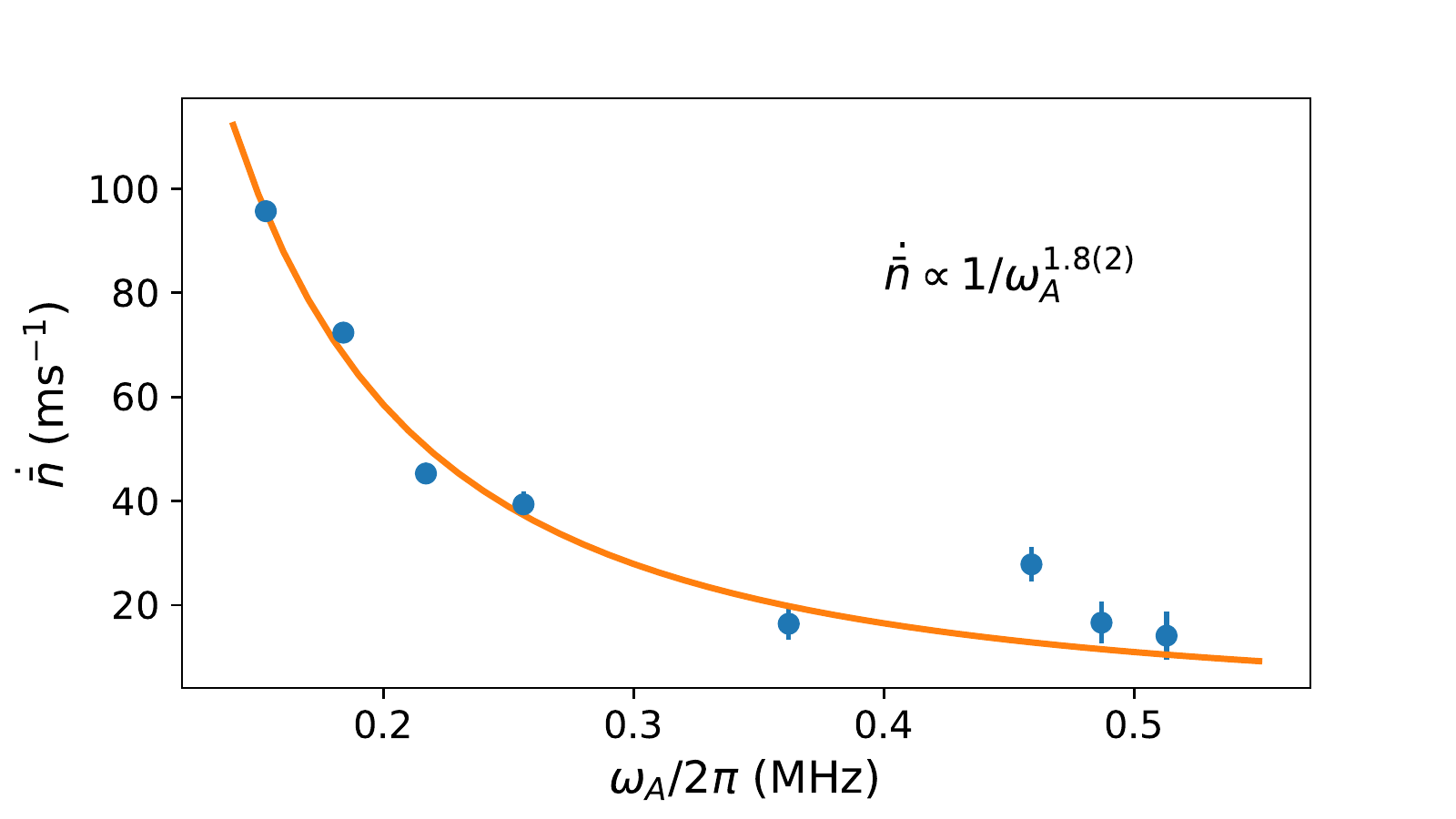}
    \caption{Extracted heating rates $\dot{\bar{n}}$ for 
    the axial mode of a single ion for eight different 
    axial frequencies are shown as the blue circles. 
    The orange solid line is the best fitted function of the
    form $\dot{\bar{n}} = c / \omega_A^\alpha$.
    }
    \label{fig:heat}
\end{figure}

We note that our model can in fact serve as a convenient tool
in experiments to extract the heating rate of the axial mode 
for a single ion or for an ion chain if its COM mode heats
much faster than the rest of the modes.
% \nam{@Ming, please fill in here with the procedure to extract
% the $\dot{\bar{n}}$.}
To obtain an accurate estimate, the static and optimized
$P_{\uparrow}$ should be measured at different delay time $\Delta t$
along with the optimal Rabi rate $\Omega_0^{\rm opt}$.
We can then fit the experimental $P_{\uparrow}$ as well as the 
ratio $\Omega_0^{\mathrm{opt}}/\Omega_0^{\mathrm{st}}$
to the theoretical predictions by adjusting the
initial temperature $\bar{n}_0$ and the heating rate $\dot{\bar{n}}$ 
as fitting parameters. In our case, we fix the initial
temperature $\bar{n}_0$ to the corresponding Doppler limit
to reduce the number of fitting parameters.
To account for all other mechanisms of
decoherence that do not depend on the motional temperature but
results in a reduction in $P_{\uparrow}$,
we include an additional fitting parameter $\delta P_{\uparrow}$ so that
the final form of the fitting functions are given by
\begin{align}
    P_{\uparrow}^{\mathrm{Exp}}(\Delta t) &\Longleftrightarrow
     P_{\uparrow}^{\mathrm{Sim}}(\bar{n}_0 + \dot{\bar{n}}\Delta t) 
     - \delta P_{\uparrow} \;, \nonumber \\
    \Omega_0^{\mathrm{opt, Exp}}(\Delta t) &\Longleftrightarrow
     \Omega_0^{\mathrm{opt, Sim}}(\bar{n}_0 + \dot{\bar{n}}\Delta t) 
     \;, \nonumber \\
    \Omega_0^{\mathrm{st, Exp}} &\Longleftrightarrow 
     \Omega_0^{\mathrm{st, Sim}}
     = \Omega_0^{\mathrm{opt, Sim}}(\bar{n}_0) \;.
\end{align}
We put this method to test by repeating the same experiment
on a single ion with axial mode frequency $\omega_A$ adjusted
to several higher values from $2\pi\times184~$kHz up to 
$2\pi\times513~$kHz.
Using the fitting method described above, we extract $\dot{\bar{n}}$
as a function of the axial mode frequency, shown in Fig.~\ref{fig:heat}.
% \nam{@Ming, discuss the figure, including the fit.}
We fit the heating rate to an inverse power law of the mode 
frequency and obtain $\dot{\bar{n}}\propto \omega_A^{-1.8(2)}$.

Our method of measuring mode temperature for a single ion complements 
the method using sideband spectroscopy, in the way that, while sideband spectroscopy
works for modes with mode vector projection along the beam propagation
direction, our method works for mode with mode vector projection
perpendicular to the beam propagation direction.
We further note that, while the examples we show are for relatively large phonon numbers,
it is straightforward to extend our method to lower phonon numbers.
This may be achieved by reducing state-preparation and measurement error and single qubit gate error, as well
as increasing $\eta$ through reducing the effective beam waist $w_x^{\rm eff}$.
Narrow band 
%\nam{Added ``Narrow band''} 
composite pulse sequences that amplify amplitude errors in
the qubit space can also be employed to increase the sensitivity of $P_{\uparrow}$
to the heating rate.
Note for an ion chain with more than one ion, the sensitivity is reduced
due to the fact that $\eta$ is generally proportional to $1/\sqrt{N}$.
%Also one needs to be careful interpreting the data when more
%than one mode heats significantly.
% \nam{No need to say this.}

\subsection{Compensating pulse sequences}

\begin{figure}
    \centering
    \includegraphics[width=1.05\columnwidth]{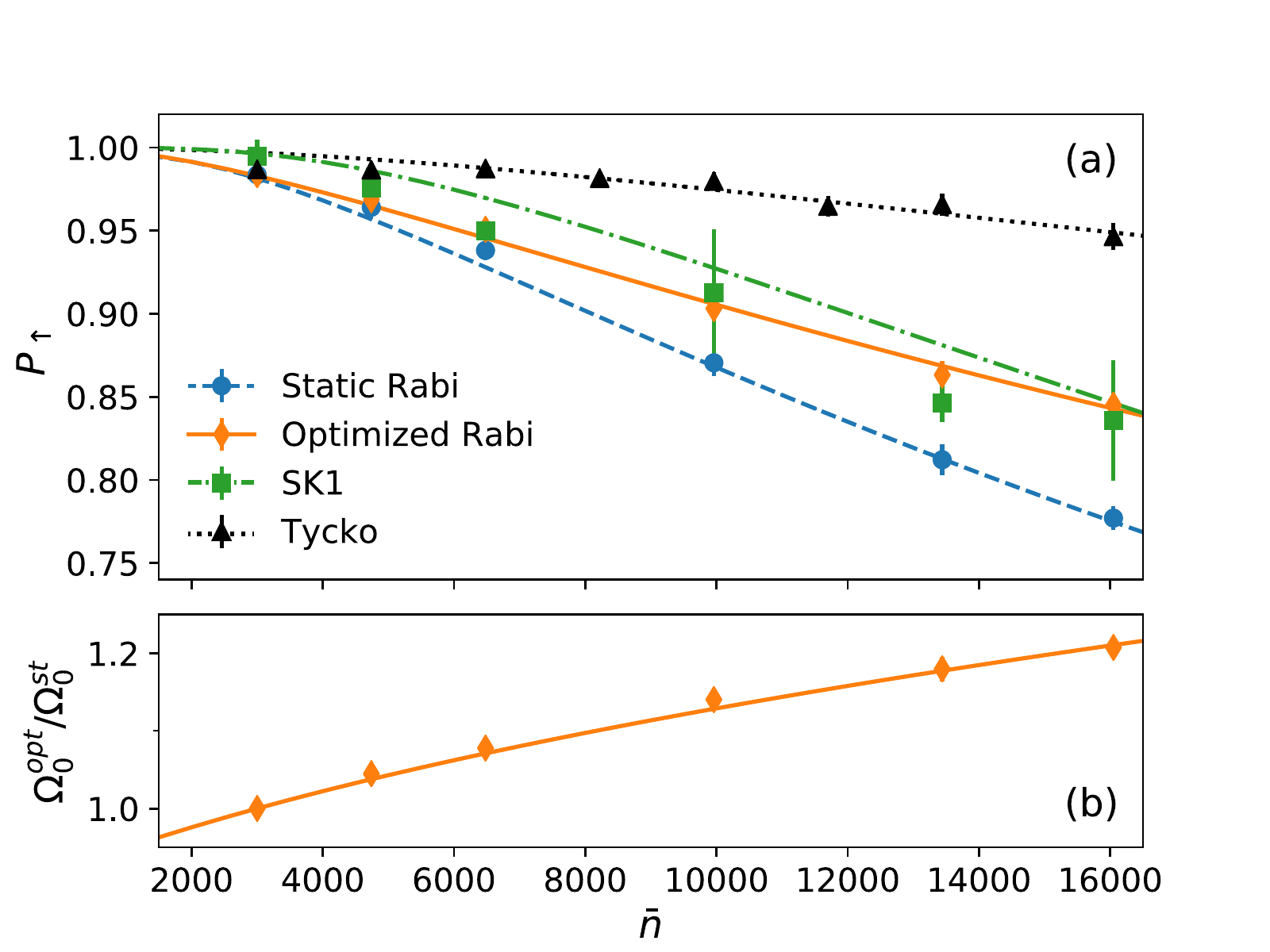}
    \caption{Axial-heating error mitigation by compensating pulse sequences.
    We implement the delayed gate protocol (A)-(D) detailed in main text,
    while also using the SK1 and the Tycko three-pulse sequences.
    A 25-ion chain with the axial COM mode frequency $\omega_A = 2\pi\times148~$kHz
    was used in a counter-propagating set up. (a) $P_{\uparrow}$ as a function
    of $\bar{n}$. Experimental (points) and simulation 
    (lines) results show the Tycko three-pulse (black triangles and dotted line) and 
    SK1 (green squares and dash-dot line) compensating sequences
    effectively mitigate the $P_{\uparrow}$ decay induced by the mode heating,
    observed by their better performance over the optimized (orange diamonds and solid line) and
    static (blue circles and dashed line) Rabi approaches without the compensating sequences.
    The simulation results for the SK1 or the Tycko three-pulse
    sequences assume a systematic phase error of $0.4$ radian per gate.
    (b) the ratio $\Omega_0^{\rm opt}/\Omega_0^{\rm st}$ 
    as a function of $\bar{n}$. The orange diamonds are the experimental 
    results and the orange solid line is the theoretical simulation results.
%    \ming{Maybe too much description.. Can we simplify this?}
}
    \label{fig:25ion148}
\end{figure}

% It is a common practice in quantum computing to employ
% pulse sequencing techniques \cite{} that mitigate certain
% quantum computational errors. 
The error induced by excessive temperature of an axial thermal mode 
is in essence an amplitude error in the quantum gate unitary,
which can be mitigated by composite pulse sequencing techniques~\cite{PulseSeqRMP,PulseSeq}
that are designed to target amplitude errors.
In this section, we experimentally demonstrate the efficacy of the well-known 
SK1 pulse sequence and the Tycko three-pulse sequence~\cite{RTycko83,PulseSeqRMP}
(see below for detail) in mitigating the axial-temperature driven error.
% \ming{Need to eventually find out what this ``BB1'' pulse is that we use.}
 %in Fig.~\ref{fig:25ion148}. 
%\ming{Maybe we need to spell out the names SK1 and BB1? Or find a more
%accurate or descriptive names for them?}
Specifically, we measure the bright populations $P_{\uparrow}$ 
as a function of $\bar{n}$ as done previously, along with simulations.

A single-qubit gate $\mathcal{R}(\theta, \phi)$ 
that rotates a Bloch vector by $\theta$
about the rotation axis on the equator of the Bloch sphere with 
polar angle $\phi$ may be parametrized as
\begin{equation}
    \mathcal{R}(\theta, \phi) = 
    \begin{pmatrix}
      \cos\frac{\theta}{2} & -ie^{-i\phi}\sin\frac{\theta}{2} \\
      -ie^{i\phi}\sin\frac{\theta}{2} & \cos\frac{\theta}{2}
    \end{pmatrix} \;.
\end{equation}
Then, the SK1 pulse sequence $\mathcal{R}_{\rm SK1} (\pi,0)$ is given by
\begin{equation}
\mathcal{R}_{\rm SK1} (\pi,0) = \mathcal{R}(2\pi,-\psi)\mathcal{R}(2\pi,\psi)\mathcal{R}(\pi,0),
\label{eq:SK1}
\end{equation}
where $\psi = \arccos(-1/4)$. 
The Tycko three-pulse sequence $\mathcal{R}_{\rm Tycko} (\pi,0)$ is given by
\begin{equation}
\mathcal{R}_{\rm Tycko} (\pi,0) = \mathcal{R}(\pi,2\pi/3) \mathcal{R}(\pi,4\pi/3) \mathcal{R}(\pi,2\pi/3).
\end{equation}
In practice, we implement
$\mathcal{R}(2\pi,\psi)$ in (\ref{eq:SK1}) by executing $\mathcal{R}(\pi,\psi)$ twice,
and similarly for $\mathcal{R}(2\pi,-\psi)$.

Figure~\ref{fig:25ion148} shows the bright population $P_{\uparrow}$
as a function of $\bar{n}$ for the static and optimized Rabi rates,
in addition to the SK1 and the Tycko three-pulse sequences introduced above.
For the sequence-based approaches, we used 
the optimal Rabi rate $\Omega_0^{\rm opt}$ approach detailed in Sec.~\ref{sec:Improve}.
The experiment here was performed on the middle ion of a 25-ion chain
which has a COM mode with a mode frequency $\omega_A = 2\pi\times148~$kHz
that heats the fastest. A counter-propagating set up is used 
where the individually addressing narrowly focused beam has a waist 
of $0.87(2)~\mu$m along the $x$-axis, while the globally addressing beam 
has a waist of $\sim 200~\mu$m. Sideband cooling of the horizontal modes
is implemented before state preparation.

%\nam{Jason's comment: In Fig 4. Really only the Tycko looks better than optimized, for the data at least (though both composite pulse sequences look better than static Rabi).  I know this is due to light shifts being uncalibrated here, and you do mention that later although maybe it's worth being clearer on that here?  Also, I guess it's also further stressed in Fig 5, where light shifts aren't included.}
From Fig.~\ref{fig:25ion148}, it is evident 
that these pulse sequences provide further improvement
in quantum gate fidelity compared to the optimized Rabi approach, 
with the Tycko three-pulse sequence especially standing out.
Note to reach the agreement between the experimental data and the simulation result for the pulse sequences,
we assumed a progressively increasing phase error of $0.4$ radian per gate 
that can be attributed to miscalibrated light shift
as well as the qubit frequency error. Such a phase error
has no effect on the $P_{\uparrow}$ of a single Rabi pulse, 
but reduces the efficacy of the composite pulse sequences.
A better calibrated single qubit gate as the basis gate
that composes the SK1 or the Tycko three-pulse sequences will 
improve the $P_{\uparrow}$ even further.
%We briefly discuss the improvement of quantum gate fidelity
%using these pulse sequences in the high fidelity regime in
%Sec.~\ref{sec:discussion}.

%\ming{@Nam: just added this. please check.}
%\nam{@Ming: This paragraph is superfluous and is redundant (see the end of the last subsection). 
\comment{\nam{If any referee asks we'll explain it in the response. I suggest we delete.}
The reason that the $P_{\uparrow}$ decays much slower with respect 
to $\bar{n}$ comparing to the single ion case is that the $c$ coefficients
for a 25-ion chain is a factor of $\sqrt{N} = 5$ smaller than for a
single ion, which effectively makes $\eta$ a factor of $5$ smaller as well.
However, albeit the apparent gain in this respect, the 25-ion chain also
heats much faster comparing to a single ion in terms of total kinetic energy
due to the simple fact that more ions are exposed to environmental noises
that heats the ion motion.
}

\section{Discussion}
\label{sec:discussion}

\begin{figure}
    \centering
    \includegraphics[width=1.05\columnwidth]{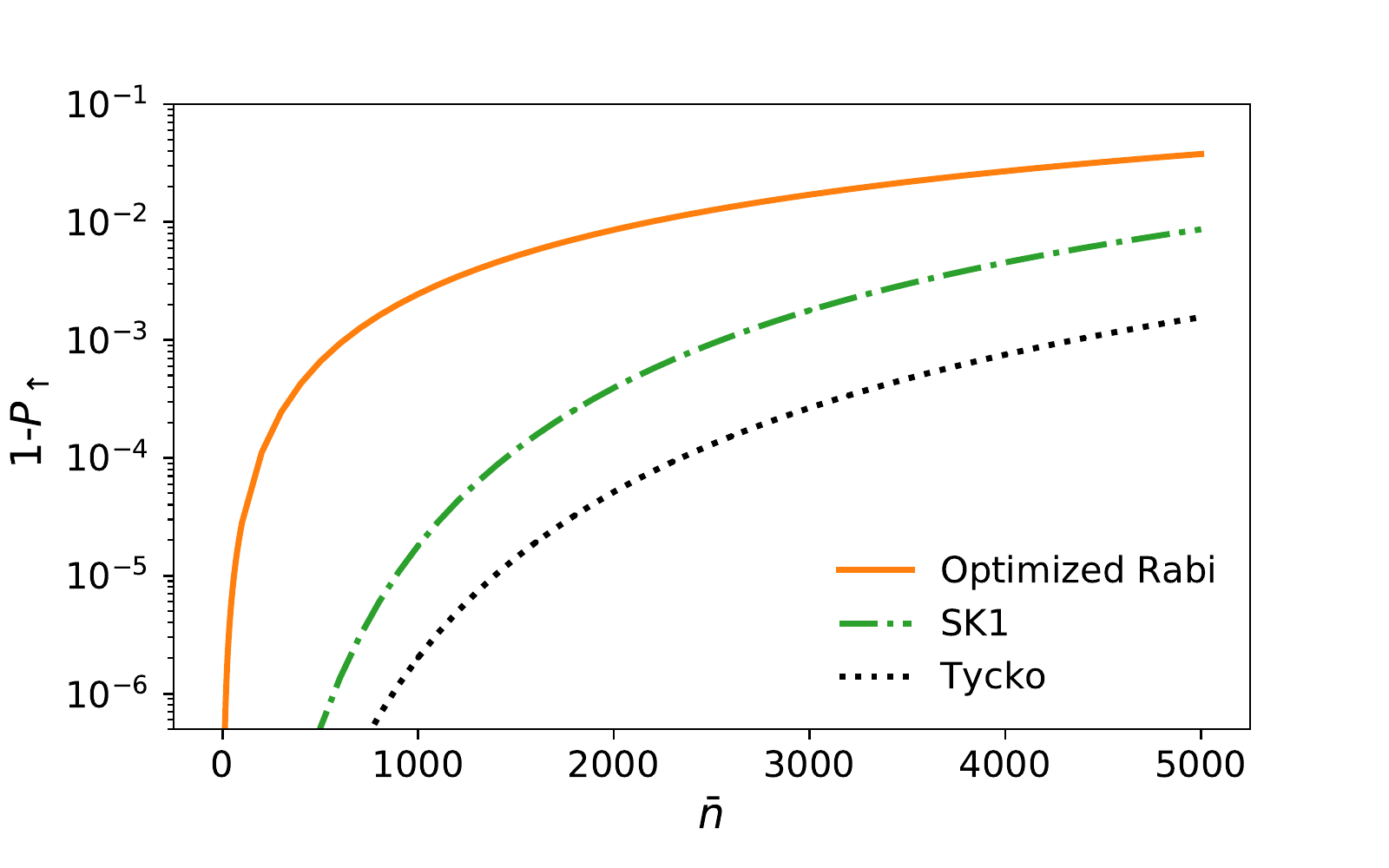}
    \caption{Theoretical simulation of the infidelity of the
    final state, represented by $1-P_{\uparrow}$, of
    the middle ion in a 25-ion chain as a function of $\bar{n}$.
    Only the decoherence effect of the axial COM mode with a frequency of 
    $2\pi\times148$ kHz is considered.
    The orange solid line represents the optimized Rabi approach,
    the green dash-dotted line represents the SK1 pulse
    sequence case, and the black dotted line represents the
    Tycko three-pulse sequence case.}
    \label{fig:25iondetail}
\end{figure}

In this paper, we have investigated the effect of non-idealities in
ion-beam geometry in quantum gate fidelity. 
% that would be pronounced in a long ion chain of ions. 
Specifically, we have provided a general theoretical framework 
that can be used to systematically examine the effect of
the coupling between the external degrees of freedom of the 
ion-light field system and the qubit space on quantum gate
operations to any desirable accuracy.
As a concrete, explicit example, we performed
a comprehensive analysis focused on the effect of excessive
axial-mode temperature and confirmed our model's validity
by comparison with experiments. Guided by our model,
we successfully mitigated the effect and increased our quantum
gate fidelity.

To further improve, we suggest the following.
From our model, it is straightforward to show that 
the gate fidelity improves rapidly when decreasing the temperature
(see Fig.~\ref{fig:25iondetail}). 
Therefore, from the hardware design point of view,
reducing heating rate of a trap itself a modest amount would
help dramatically. We can also consider efficient sympathetic cooling scheme 
during a quantum circuit execution as well~\cite{symcool1,symcool2}, keeping $\bar{n}$ to an acceptable
level throughout the quantum computational runtime. 
Raising axial mode frequencies by either decreasing
the ion spacing or using optical tweezers~\cite{YCShen20}
is also a viable way to improve quantum gate fidelity, since this
helps reduce the size of $\eta$ and may reduce the heating rate. 
We further note that increasing the waist of the 
individually addressing beam will directly decrease $\eta$, thus
reducing the undesirable decoherence. Finally, composite pulse sequences,
as we have demonstrated, can significantly improve fidelity.
%\nam{It is designed to significantly improve the gate fidelity in the
%high fidelity regime so no surprise here.}
Figure~\ref{fig:25iondetail} shows additional simulation data
that shows the expected infidelity $1-P_{\uparrow}$ for the SK1 and the Tycko three-pulse sequences.
It is clear that  
the composite pulse sequences can significantly increase the range
of $\bar{n}$ acceptable for a successful quantum gate operation with
high fidelity.
% Of course, one must always be aware of trade-offs brought about by 
% making these changes.  

Looking forward, building on our exercise,
additional terms in the interaction Hamiltonian can now be systematically included
in descending order of their contribution towards quantum gate infidelity
to help achieve high fidelity trapped-ion quantum computing.
For example, we can include the Debye-Waller
effect inducing terms $B_0^\pm$. We can consider higher order
terms in the $B_1^{\pm}$ function as well that originated from the Gouy phase.
Note the latter will manifest themselves as a small correction to the
Debye-Waller effect. These terms will induce decoherence, 
if the temperature of the motional modes of an ion chain is high
and/or the misalignment between the ion and its addressing beam is large. 
Our framework analytically captures these effects accurately
and provides quantitative methodologies to characterize their
impact on quantum gate fidelity.

Although we have largely focused on the effect on single-qubit 
gate operations in our analysis and experiments, pertaining 
the coupling to the axial modes, similar derivation and 
analysis can readily be extended to two-qubit gates. In fact, 
most of the conclusion including mitigation
strategies and techniques for single-qubit gates hold analogous
and similar counterparts for two-qubit gates.

% refer to specific sections.
% We confirm the validity of our model by a carefully designed experiment,
% whose results agree well with our expectations, as demonstrated in
% Sec.~\ref{}. In light of our discovery, we can provide concrete suggestions
% as to how the trapped-ion quantum computing platform may be able to scale
% as follows.

%From the analysis of the effect of excessive axial mode temperature,
%we have already laid out a few ways to improve quantum gate operations.

% \begin{itemize}
% % \item implications to QC -- transverse temperature matters for high fidelity computation. can we make some scaling arguments about radial frequency, $1/\omega^3$ heating rate, ion spacing to limit where this error dominates over radial heating error, implying a max chain length?
% \item Discuss possibilities of further error analysis with our framework.
% \item Discuss two-qubit gates.
% \item Mitigation: tweezers array to increase axial frequency, sympathetic cooling, pulse shaping techniques.
% \item Light shift and other effects from the eliminated atomic levels.
% \item Other higher order effects: dipole force, Doppler shift, anharmonicity, etc.
% \end{itemize}

\section{Conclusion}
\label{sec:conclusion}
In this paper, we have derived a general Hamiltonian
capable of pinpointing the sources of infidelity in
trapped-ion quantum computers with a long chain.
By carefully analyzing the Hamiltonian with realistic beam geometry 
and parameters, quantum computational errors incurred due to
alignment and focus have been identified and experiments were
conducted to confirm their existence.
Our framework is versatile, precisely laying out 
all terms of importance according to the quality requirement
for any trapped-ion quantum computing platform. 
We expect our results will help guide the quantum
hardware engineers to make informed decisions,
tailored for future hardware design criteria.

\section{Acknowledgement}

The authors would like to thank Shantanu Debnath and Kevin Landsman for their contribution in designing and implementing experimental system components and
Christopher Monroe, Jungsang Kim, and Marco Cetina for helpful discussions.

\bibliography{axial}

\begin{thebibliography}{30}%
\makeatletter
\providecommand \@ifxundefined [1]{%
 \@ifx{#1\undefined}
}%
\providecommand \@ifnum [1]{%
 \ifnum #1\expandafter \@firstoftwo
 \else \expandafter \@secondoftwo
 \fi
}%
\providecommand \@ifx [1]{%
 \ifx #1\expandafter \@firstoftwo
 \else \expandafter \@secondoftwo
 \fi
}%
\providecommand \natexlab [1]{#1}%
\providecommand \enquote  [1]{``#1''}%
\providecommand \bibnamefont  [1]{#1}%
\providecommand \bibfnamefont [1]{#1}%
\providecommand \citenamefont [1]{#1}%
\providecommand \href@noop [0]{\@secondoftwo}%
\providecommand \href [0]{\begingroup \@sanitize@url \@href}%
\providecommand \@href[1]{\@@startlink{#1}\@@href}%
\providecommand \@@href[1]{\endgroup#1\@@endlink}%
\providecommand \@sanitize@url [0]{\catcode `\\12\catcode `\$12\catcode
  `\&12\catcode `\#12\catcode `\^12\catcode `\_12\catcode `\%12\relax}%
\providecommand \@@startlink[1]{}%
\providecommand \@@endlink[0]{}%
\providecommand \url  [0]{\begingroup\@sanitize@url \@url }%
\providecommand \@url [1]{\endgroup\@href {#1}{\urlprefix }}%
\providecommand \urlprefix  [0]{URL }%
\providecommand \Eprint [0]{\href }%
\providecommand \doibase [0]{http://dx.doi.org/}%
\providecommand \selectlanguage [0]{\@gobble}%
\providecommand \bibinfo  [0]{\@secondoftwo}%
\providecommand \bibfield  [0]{\@secondoftwo}%
\providecommand \translation [1]{[#1]}%
\providecommand \BibitemOpen [0]{}%
\providecommand \bibitemStop [0]{}%
\providecommand \bibitemNoStop [0]{.\EOS\space}%
\providecommand \EOS [0]{\spacefactor3000\relax}%
\providecommand \BibitemShut  [1]{\csname bibitem#1\endcsname}%
\let\auto@bib@innerbib\@empty
%</preamble>
\bibitem [{\citenamefont {Gaebler}\ \emph {et~al.}(2016)\citenamefont
  {Gaebler}, \citenamefont {Tan}, \citenamefont {Lin}, \citenamefont {Wan},
  \citenamefont {Bowler}, \citenamefont {Keith}, \citenamefont {Glancy},
  \citenamefont {Coakley}, \citenamefont {Knill}, \citenamefont {Leibfried},\
  and\ \citenamefont {Wineland}}]{JGaebler16}%
  \BibitemOpen
  \bibfield  {author} {\bibinfo {author} {\bibfnamefont {J.~P.}\ \bibnamefont
  {Gaebler}}, \bibinfo {author} {\bibfnamefont {T.~R.}\ \bibnamefont {Tan}},
  \bibinfo {author} {\bibfnamefont {Y.}~\bibnamefont {Lin}}, \bibinfo {author}
  {\bibfnamefont {Y.}~\bibnamefont {Wan}}, \bibinfo {author} {\bibfnamefont
  {R.}~\bibnamefont {Bowler}}, \bibinfo {author} {\bibfnamefont {A.~C.}\
  \bibnamefont {Keith}}, \bibinfo {author} {\bibfnamefont {S.}~\bibnamefont
  {Glancy}}, \bibinfo {author} {\bibfnamefont {K.}~\bibnamefont {Coakley}},
  \bibinfo {author} {\bibfnamefont {E.}~\bibnamefont {Knill}}, \bibinfo
  {author} {\bibfnamefont {D.}~\bibnamefont {Leibfried}}, \ and\ \bibinfo
  {author} {\bibfnamefont {D.~J.}\ \bibnamefont {Wineland}},\ }\href {\doibase
  10.1103/PhysRevLett.117.060505} {\bibfield  {journal} {\bibinfo  {journal}
  {Phys. Rev. Lett.}\ }\textbf {\bibinfo {volume} {117}},\ \bibinfo {pages}
  {060505} (\bibinfo {year} {2016})}\BibitemShut {NoStop}%
\bibitem [{\citenamefont {Ballance}\ \emph {et~al.}(2016)\citenamefont
  {Ballance}, \citenamefont {Harty}, \citenamefont {Linke}, \citenamefont
  {Sepiol},\ and\ \citenamefont {Lucas}}]{CBallance16}%
  \BibitemOpen
  \bibfield  {author} {\bibinfo {author} {\bibfnamefont {C.~J.}\ \bibnamefont
  {Ballance}}, \bibinfo {author} {\bibfnamefont {T.~P.}\ \bibnamefont {Harty}},
  \bibinfo {author} {\bibfnamefont {N.~M.}\ \bibnamefont {Linke}}, \bibinfo
  {author} {\bibfnamefont {M.~A.}\ \bibnamefont {Sepiol}}, \ and\ \bibinfo
  {author} {\bibfnamefont {D.~M.}\ \bibnamefont {Lucas}},\ }\href {\doibase
  10.1103/PhysRevLett.117.060504} {\bibfield  {journal} {\bibinfo  {journal}
  {Phys. Rev. Lett.}\ }\textbf {\bibinfo {volume} {117}},\ \bibinfo {pages}
  {060504} (\bibinfo {year} {2016})}\BibitemShut {NoStop}%
\bibitem [{\citenamefont {Kielpinski}\ \emph {et~al.}(2002)\citenamefont
  {Kielpinski}, \citenamefont {Monroe},\ and\ \citenamefont {Wineland}}]{qccd}%
  \BibitemOpen
  \bibfield  {author} {\bibinfo {author} {\bibfnamefont {D.}~\bibnamefont
  {Kielpinski}}, \bibinfo {author} {\bibfnamefont {C.}~\bibnamefont {Monroe}},
  \ and\ \bibinfo {author} {\bibfnamefont {D.~J.}\ \bibnamefont {Wineland}},\
  }\href {\doibase 10.1038/nature00784} {\bibfield  {journal} {\bibinfo
  {journal} {Nature}\ }\textbf {\bibinfo {volume} {417}},\ \bibinfo {pages}
  {709} (\bibinfo {year} {2002})}\BibitemShut {NoStop}%
\bibitem [{\citenamefont {Pino}\ \emph {et~al.}(2020)\citenamefont {Pino},
  \citenamefont {Dreiling}, \citenamefont {Figgatt}, \citenamefont {Gaebler},
  \citenamefont {Moses}, \citenamefont {Allman}, \citenamefont {Baldwin},
  \citenamefont {Foss-Feig}, \citenamefont {Hayes}, \citenamefont {Mayer},
  \citenamefont {Ryan-Anderson},\ and\ \citenamefont
  {Neyenhuis}}]{qccdHoneywell}%
  \BibitemOpen
  \bibfield  {author} {\bibinfo {author} {\bibfnamefont {J.~M.}\ \bibnamefont
  {Pino}}, \bibinfo {author} {\bibfnamefont {J.}~\bibnamefont {Dreiling}},
  \bibinfo {author} {\bibfnamefont {C.}~\bibnamefont {Figgatt}}, \bibinfo
  {author} {\bibfnamefont {J.}~\bibnamefont {Gaebler}}, \bibinfo {author}
  {\bibfnamefont {S.}~\bibnamefont {Moses}}, \bibinfo {author} {\bibfnamefont
  {M.~S.}\ \bibnamefont {Allman}}, \bibinfo {author} {\bibfnamefont
  {C.}~\bibnamefont {Baldwin}}, \bibinfo {author} {\bibfnamefont
  {M.}~\bibnamefont {Foss-Feig}}, \bibinfo {author} {\bibfnamefont
  {D.}~\bibnamefont {Hayes}}, \bibinfo {author} {\bibfnamefont
  {K.}~\bibnamefont {Mayer}}, \bibinfo {author} {\bibfnamefont
  {C.}~\bibnamefont {Ryan-Anderson}}, \ and\ \bibinfo {author} {\bibfnamefont
  {B.}~\bibnamefont {Neyenhuis}},\ }\href@noop {} {\bibfield  {journal}
  {\bibinfo  {journal} {arXiv preprint arXiv:2003.01293}\ } (\bibinfo {year}
  {2020})}\BibitemShut {NoStop}%
\bibitem [{\citenamefont {Monroe}\ \emph {et~al.}(2014)\citenamefont {Monroe},
  \citenamefont {Raussendorf}, \citenamefont {Ruthven}, \citenamefont {Brown},
  \citenamefont {Maunz}, \citenamefont {Duan},\ and\ \citenamefont
  {Kim}}]{photonicInterconnect}%
  \BibitemOpen
  \bibfield  {author} {\bibinfo {author} {\bibfnamefont {C.}~\bibnamefont
  {Monroe}}, \bibinfo {author} {\bibfnamefont {R.}~\bibnamefont {Raussendorf}},
  \bibinfo {author} {\bibfnamefont {A.}~\bibnamefont {Ruthven}}, \bibinfo
  {author} {\bibfnamefont {K.~R.}\ \bibnamefont {Brown}}, \bibinfo {author}
  {\bibfnamefont {P.}~\bibnamefont {Maunz}}, \bibinfo {author} {\bibfnamefont
  {L.-M.}\ \bibnamefont {Duan}}, \ and\ \bibinfo {author} {\bibfnamefont
  {J.}~\bibnamefont {Kim}},\ }\href {\doibase 10.1103/PhysRevA.89.022317}
  {\bibfield  {journal} {\bibinfo  {journal} {Phys. Rev. A}\ }\textbf {\bibinfo
  {volume} {89}},\ \bibinfo {pages} {022317} (\bibinfo {year}
  {2014})}\BibitemShut {NoStop}%
\bibitem [{\citenamefont {Linke}\ \emph {et~al.}(2017)\citenamefont {Linke},
  \citenamefont {Maslov}, \citenamefont {Roetteler}, \citenamefont {Debnath},
  \citenamefont {Figgatt}, \citenamefont {Landsman}, \citenamefont {Wright},\
  and\ \citenamefont {Monroe}}]{Linke_2017}%
  \BibitemOpen
  \bibfield  {author} {\bibinfo {author} {\bibfnamefont {N.~M.}\ \bibnamefont
  {Linke}}, \bibinfo {author} {\bibfnamefont {D.}~\bibnamefont {Maslov}},
  \bibinfo {author} {\bibfnamefont {M.}~\bibnamefont {Roetteler}}, \bibinfo
  {author} {\bibfnamefont {S.}~\bibnamefont {Debnath}}, \bibinfo {author}
  {\bibfnamefont {C.}~\bibnamefont {Figgatt}}, \bibinfo {author} {\bibfnamefont
  {K.~A.}\ \bibnamefont {Landsman}}, \bibinfo {author} {\bibfnamefont
  {K.}~\bibnamefont {Wright}}, \ and\ \bibinfo {author} {\bibfnamefont
  {C.}~\bibnamefont {Monroe}},\ }\href {\doibase 10.1073/pnas.1618020114}
  {\bibfield  {journal} {\bibinfo  {journal} {Proceedings of the National
  Academy of Sciences}\ }\textbf {\bibinfo {volume} {114}},\ \bibinfo {pages}
  {3305–3310} (\bibinfo {year} {2017})}\BibitemShut {NoStop}%
\bibitem [{\citenamefont {Nam}\ and\ \citenamefont {Maslov}(2019)}]{LowCost}%
  \BibitemOpen
  \bibfield  {author} {\bibinfo {author} {\bibfnamefont {Y.}~\bibnamefont
  {Nam}}\ and\ \bibinfo {author} {\bibfnamefont {D.}~\bibnamefont {Maslov}},\
  }\href {\doibase 10.1038/s41534-019-0152-0} {\bibfield  {journal} {\bibinfo
  {journal} {npj Quantum Information}\ }\textbf {\bibinfo {volume} {5}}
  (\bibinfo {year} {2019}),\ 10.1038/s41534-019-0152-0}\BibitemShut {NoStop}%
\bibitem [{\citenamefont {{Maslov}}\ \emph {et~al.}(2019)\citenamefont
  {{Maslov}}, \citenamefont {{Nam}},\ and\ \citenamefont {{Kim}}}]{IEEE}%
  \BibitemOpen
  \bibfield  {author} {\bibinfo {author} {\bibfnamefont {D.}~\bibnamefont
  {{Maslov}}}, \bibinfo {author} {\bibfnamefont {Y.}~\bibnamefont {{Nam}}}, \
  and\ \bibinfo {author} {\bibfnamefont {J.}~\bibnamefont {{Kim}}},\
  }\href@noop {} {\bibfield  {journal} {\bibinfo  {journal} {Proceedings of the
  IEEE}\ }\textbf {\bibinfo {volume} {107}},\ \bibinfo {pages} {5} (\bibinfo
  {year} {2019})}\BibitemShut {NoStop}%
\bibitem [{\citenamefont {Grzesiak}\ \emph {et~al.}(2020)\citenamefont
  {Grzesiak}, \citenamefont {Blümel}, \citenamefont {Wright}, \citenamefont
  {Beck}, \citenamefont {Pisenti}, \citenamefont {Li}, \citenamefont {Chaplin},
  \citenamefont {Amini}, \citenamefont {Debnath}, \citenamefont {Chen},\ and\
  \citenamefont {et~al.}}]{EASE}%
  \BibitemOpen
  \bibfield  {author} {\bibinfo {author} {\bibfnamefont {N.}~\bibnamefont
  {Grzesiak}}, \bibinfo {author} {\bibfnamefont {R.}~\bibnamefont {Blümel}},
  \bibinfo {author} {\bibfnamefont {K.}~\bibnamefont {Wright}}, \bibinfo
  {author} {\bibfnamefont {K.~M.}\ \bibnamefont {Beck}}, \bibinfo {author}
  {\bibfnamefont {N.~C.}\ \bibnamefont {Pisenti}}, \bibinfo {author}
  {\bibfnamefont {M.}~\bibnamefont {Li}}, \bibinfo {author} {\bibfnamefont
  {V.}~\bibnamefont {Chaplin}}, \bibinfo {author} {\bibfnamefont {J.~M.}\
  \bibnamefont {Amini}}, \bibinfo {author} {\bibfnamefont {S.}~\bibnamefont
  {Debnath}}, \bibinfo {author} {\bibfnamefont {J.-S.}\ \bibnamefont {Chen}}, \
  and\ \bibinfo {author} {\bibnamefont {et~al.}},\ }\href {\doibase
  10.1038/s41467-020-16790-9} {\bibfield  {journal} {\bibinfo  {journal}
  {Nature Communications}\ }\textbf {\bibinfo {volume} {11}} (\bibinfo {year}
  {2020}),\ 10.1038/s41467-020-16790-9}\BibitemShut {NoStop}%
\bibitem [{\citenamefont {Vandersypen}\ and\ \citenamefont
  {Chuang}(2005)}]{PulseSeqRMP}%
  \BibitemOpen
  \bibfield  {author} {\bibinfo {author} {\bibfnamefont {L.~M.~K.}\
  \bibnamefont {Vandersypen}}\ and\ \bibinfo {author} {\bibfnamefont {I.~L.}\
  \bibnamefont {Chuang}},\ }\href {\doibase 10.1103/RevModPhys.76.1037}
  {\bibfield  {journal} {\bibinfo  {journal} {Rev. Mod. Phys.}\ }\textbf
  {\bibinfo {volume} {76}},\ \bibinfo {pages} {1037} (\bibinfo {year}
  {2005})}\BibitemShut {NoStop}%
\bibitem [{\citenamefont {Merrill}\ and\ \citenamefont
  {Brown}(2014)}]{PulseSeq}%
  \BibitemOpen
  \bibfield  {author} {\bibinfo {author} {\bibfnamefont {J.~T.}\ \bibnamefont
  {Merrill}}\ and\ \bibinfo {author} {\bibfnamefont {K.~R.}\ \bibnamefont
  {Brown}},\ }\enquote {\bibinfo {title} {Progress in compensating pulse
  sequences for quantum computation},}\ in\ \href {\doibase
  10.1002/9781118742631.ch10} {\emph {\bibinfo {booktitle} {Quantum Information
  and Computation for Chemistry}}}\ (\bibinfo  {publisher} {John Wiley \& Sons,
  Ltd},\ \bibinfo {year} {2014})\ pp.\ \bibinfo {pages} {241--294}\BibitemShut
  {NoStop}%
\bibitem [{\citenamefont {Cetina}\ \emph {et~al.}(2020)\citenamefont {Cetina},
  \citenamefont {Egan}, \citenamefont {Noel}, \citenamefont {Goldman},
  \citenamefont {Risinger}, \citenamefont {Zhu}, \citenamefont {Biswas},\ and\
  \citenamefont {Monroe}}]{marko}%
  \BibitemOpen
  \bibfield  {author} {\bibinfo {author} {\bibfnamefont {M.}~\bibnamefont
  {Cetina}}, \bibinfo {author} {\bibfnamefont {L.}~\bibnamefont {Egan}},
  \bibinfo {author} {\bibfnamefont {C.~A.}\ \bibnamefont {Noel}}, \bibinfo
  {author} {\bibfnamefont {M.~L.}\ \bibnamefont {Goldman}}, \bibinfo {author}
  {\bibfnamefont {A.~R.}\ \bibnamefont {Risinger}}, \bibinfo {author}
  {\bibfnamefont {D.}~\bibnamefont {Zhu}}, \bibinfo {author} {\bibfnamefont
  {D.}~\bibnamefont {Biswas}}, \ and\ \bibinfo {author} {\bibfnamefont
  {C.}~\bibnamefont {Monroe}},\ }\href@noop {} {\bibfield  {journal} {\bibinfo
  {journal} {arXiv preprint arXiv:2007.06768}\ } (\bibinfo {year}
  {2020})}\BibitemShut {NoStop}%
\bibitem [{\citenamefont {West}\ \emph {et~al.}(2020)\citenamefont {West},
  \citenamefont {Putnam}, \citenamefont {Campbell},\ and\ \citenamefont
  {Hamilton}}]{WesAxial}%
  \BibitemOpen
  \bibfield  {author} {\bibinfo {author} {\bibfnamefont {A.}~\bibnamefont
  {West}}, \bibinfo {author} {\bibfnamefont {R.}~\bibnamefont {Putnam}},
  \bibinfo {author} {\bibfnamefont {W.}~\bibnamefont {Campbell}}, \ and\
  \bibinfo {author} {\bibfnamefont {P.}~\bibnamefont {Hamilton}},\ }\href@noop
  {} {\bibfield  {journal} {\bibinfo  {journal} {arXiv preprint
  arXiv:2007.10437}\ } (\bibinfo {year} {2020})}\BibitemShut {NoStop}%
\bibitem [{\citenamefont {Brion}\ \emph {et~al.}(2007)\citenamefont {Brion},
  \citenamefont {Pedersen},\ and\ \citenamefont {M{\o}lmer}}]{EBrion2007}%
  \BibitemOpen
  \bibfield  {author} {\bibinfo {author} {\bibfnamefont {E.}~\bibnamefont
  {Brion}}, \bibinfo {author} {\bibfnamefont {L.~H.}\ \bibnamefont {Pedersen}},
  \ and\ \bibinfo {author} {\bibfnamefont {K.}~\bibnamefont {M{\o}lmer}},\
  }\href {\doibase 10.1088/1751-8113/40/5/011} {\bibfield  {journal} {\bibinfo
  {journal} {Journal of Physics A: Mathematical and Theoretical}\ }\textbf
  {\bibinfo {volume} {40}},\ \bibinfo {pages} {1033} (\bibinfo {year}
  {2007})}\BibitemShut {NoStop}%
\bibitem [{\citenamefont {James}(1998)}]{DFVJames1998}%
  \BibitemOpen
  \bibfield  {author} {\bibinfo {author} {\bibfnamefont {D.~F.~V.}\
  \bibnamefont {James}},\ }\href {\doibase 10.1007/s003400050373} {\bibfield
  {journal} {\bibinfo  {journal} {Applied Physics B}\ }\textbf {\bibinfo
  {volume} {66}},\ \bibinfo {pages} {181} (\bibinfo {year} {1998})}\BibitemShut
  {NoStop}%
\bibitem [{\citenamefont {Siegman}(1986)}]{AESiegman1986}%
  \BibitemOpen
  \bibfield  {author} {\bibinfo {author} {\bibfnamefont {A.~E.}\ \bibnamefont
  {Siegman}},\ }\href@noop {} {\emph {\bibinfo {title} {Lasers}}}\ (\bibinfo
  {publisher} {University Science Books},\ \bibinfo {address} {Mill Valley,
  Calif.},\ \bibinfo {year} {1986})\BibitemShut {NoStop}%
\bibitem [{\citenamefont {M\o{}lmer}\ and\ \citenamefont
  {S\o{}rensen}(1999)}]{MS1}%
  \BibitemOpen
  \bibfield  {author} {\bibinfo {author} {\bibfnamefont {K.}~\bibnamefont
  {M\o{}lmer}}\ and\ \bibinfo {author} {\bibfnamefont {A.}~\bibnamefont
  {S\o{}rensen}},\ }\href {\doibase 10.1103/PhysRevLett.82.1835} {\bibfield
  {journal} {\bibinfo  {journal} {Phys. Rev. Lett.}\ }\textbf {\bibinfo
  {volume} {82}},\ \bibinfo {pages} {1835} (\bibinfo {year}
  {1999})}\BibitemShut {NoStop}%
\bibitem [{\citenamefont {S\o{}rensen}\ and\ \citenamefont
  {M\o{}lmer}(1999)}]{MS2}%
  \BibitemOpen
  \bibfield  {author} {\bibinfo {author} {\bibfnamefont {A.}~\bibnamefont
  {S\o{}rensen}}\ and\ \bibinfo {author} {\bibfnamefont {K.}~\bibnamefont
  {M\o{}lmer}},\ }\href {\doibase 10.1103/PhysRevLett.82.1971} {\bibfield
  {journal} {\bibinfo  {journal} {Phys. Rev. Lett.}\ }\textbf {\bibinfo
  {volume} {82}},\ \bibinfo {pages} {1971} (\bibinfo {year}
  {1999})}\BibitemShut {NoStop}%
\bibitem [{\citenamefont {Cirac}\ and\ \citenamefont {Zoller}(1995)}]{CZ}%
  \BibitemOpen
  \bibfield  {author} {\bibinfo {author} {\bibfnamefont {J.~I.}\ \bibnamefont
  {Cirac}}\ and\ \bibinfo {author} {\bibfnamefont {P.}~\bibnamefont {Zoller}},\
  }\href {\doibase 10.1103/PhysRevLett.74.4091} {\bibfield  {journal} {\bibinfo
   {journal} {Phys. Rev. Lett.}\ }\textbf {\bibinfo {volume} {74}},\ \bibinfo
  {pages} {4091} (\bibinfo {year} {1995})}\BibitemShut {NoStop}%
\bibitem [{\citenamefont {Wineland}\ \emph {et~al.}(1998)\citenamefont
  {Wineland}, \citenamefont {Monroe}, \citenamefont {Itano}, \citenamefont
  {Leibfried}, \citenamefont {King},\ and\ \citenamefont
  {Meekhof}}]{NISTBible}%
  \BibitemOpen
  \bibfield  {author} {\bibinfo {author} {\bibfnamefont {D.~J.}\ \bibnamefont
  {Wineland}}, \bibinfo {author} {\bibfnamefont {C.}~\bibnamefont {Monroe}},
  \bibinfo {author} {\bibfnamefont {W.~M.}\ \bibnamefont {Itano}}, \bibinfo
  {author} {\bibfnamefont {D.}~\bibnamefont {Leibfried}}, \bibinfo {author}
  {\bibfnamefont {B.~E.}\ \bibnamefont {King}}, \ and\ \bibinfo {author}
  {\bibfnamefont {D.~M.}\ \bibnamefont {Meekhof}},\ }\href@noop {} {\bibfield
  {journal} {\bibinfo  {journal} {J. Res. Natl. Inst. Stand. Technol.}\
  }\textbf {\bibinfo {volume} {103}},\ \bibinfo {pages} {259} (\bibinfo {year}
  {1998})}\BibitemShut {NoStop}%
\bibitem [{\citenamefont {Bl{\"u}mel}\ \emph {et~al.}(2019)\citenamefont
  {Bl{\"u}mel}, \citenamefont {Grzesiak},\ and\ \citenamefont {Nam}}]{AMFM}%
  \BibitemOpen
  \bibfield  {author} {\bibinfo {author} {\bibfnamefont {R.}~\bibnamefont
  {Bl{\"u}mel}}, \bibinfo {author} {\bibfnamefont {N.}~\bibnamefont
  {Grzesiak}}, \ and\ \bibinfo {author} {\bibfnamefont {Y.~S.}\ \bibnamefont
  {Nam}},\ }\href@noop {} {\bibfield  {journal} {\bibinfo  {journal} {ArXiv
  preprint arXiv:1905.09292}\ } (\bibinfo {year} {2019})}\BibitemShut {NoStop}%
\bibitem [{Note1()}]{Note1}%
  \BibitemOpen
  \bibinfo {note} {For sideband cooled horizontal modes in counter-propagating
  set up, see discussion on Debye-Waller effect in Sec.~\ref
  {sec:approx}}\BibitemShut {NoStop}%
\bibitem [{Note2()}]{Note2}%
  \BibitemOpen
  \bibinfo {note} {Note that the terms with odd $l$ in Eq.~(\ref {eq:HIprime})
  are suppressed by the alignment parameter $\xi $ due to the fact that the
  zeroth order terms of the odd Hermite polynomials are on the order of $\xi $
  instead of a constant. Therefore, the coefficients in front of the imbalanced
  terms are one $\xi $ order smaller than the balanced terms.}\BibitemShut
  {Stop}%
\bibitem [{\citenamefont {Wright}\ \emph {et~al.}(2019)\citenamefont {Wright},
  \citenamefont {Beck}, \citenamefont {Debnath}, \citenamefont {Amini},
  \citenamefont {Nam}, \citenamefont {Grzesiak}, \citenamefont {Chen},
  \citenamefont {Pisenti}, \citenamefont {Chmielewski}, \citenamefont
  {Collins}, \citenamefont {Hudek}, \citenamefont {Mizrahi}, \citenamefont
  {Wong-Campos}, \citenamefont {Allen}, \citenamefont {Apisdorf}, \citenamefont
  {Solomon}, \citenamefont {Williams}, \citenamefont {Ducore}, \citenamefont
  {Blinov}, \citenamefont {Kreikemeier}, \citenamefont {Chaplin}, \citenamefont
  {Keesan}, \citenamefont {Monroe},\ and\ \citenamefont
  {Kim}}]{benchmarkPaper}%
  \BibitemOpen
  \bibfield  {author} {\bibinfo {author} {\bibfnamefont {K.}~\bibnamefont
  {Wright}}, \bibinfo {author} {\bibfnamefont {K.~M.}\ \bibnamefont {Beck}},
  \bibinfo {author} {\bibfnamefont {S.}~\bibnamefont {Debnath}}, \bibinfo
  {author} {\bibfnamefont {J.~M.}\ \bibnamefont {Amini}}, \bibinfo {author}
  {\bibfnamefont {Y.}~\bibnamefont {Nam}}, \bibinfo {author} {\bibfnamefont
  {N.}~\bibnamefont {Grzesiak}}, \bibinfo {author} {\bibfnamefont {J.-S.}\
  \bibnamefont {Chen}}, \bibinfo {author} {\bibfnamefont {N.~C.}\ \bibnamefont
  {Pisenti}}, \bibinfo {author} {\bibfnamefont {M.}~\bibnamefont
  {Chmielewski}}, \bibinfo {author} {\bibfnamefont {C.}~\bibnamefont
  {Collins}}, \bibinfo {author} {\bibfnamefont {K.~M.}\ \bibnamefont {Hudek}},
  \bibinfo {author} {\bibfnamefont {J.}~\bibnamefont {Mizrahi}}, \bibinfo
  {author} {\bibfnamefont {J.~D.}\ \bibnamefont {Wong-Campos}}, \bibinfo
  {author} {\bibfnamefont {S.}~\bibnamefont {Allen}}, \bibinfo {author}
  {\bibfnamefont {J.}~\bibnamefont {Apisdorf}}, \bibinfo {author}
  {\bibfnamefont {P.}~\bibnamefont {Solomon}}, \bibinfo {author} {\bibfnamefont
  {M.}~\bibnamefont {Williams}}, \bibinfo {author} {\bibfnamefont {A.~M.}\
  \bibnamefont {Ducore}}, \bibinfo {author} {\bibfnamefont {A.}~\bibnamefont
  {Blinov}}, \bibinfo {author} {\bibfnamefont {S.~M.}\ \bibnamefont
  {Kreikemeier}}, \bibinfo {author} {\bibfnamefont {V.}~\bibnamefont
  {Chaplin}}, \bibinfo {author} {\bibfnamefont {M.}~\bibnamefont {Keesan}},
  \bibinfo {author} {\bibfnamefont {C.}~\bibnamefont {Monroe}}, \ and\ \bibinfo
  {author} {\bibfnamefont {J.}~\bibnamefont {Kim}},\ }\href {\doibase
  10.1038/s41467-019-13534-2} {\bibfield  {journal} {\bibinfo  {journal}
  {Nature Communications}\ }\textbf {\bibinfo {volume} {10}},\ \bibinfo {pages}
  {5464} (\bibinfo {year} {2019})}\BibitemShut {NoStop}%
\bibitem [{\citenamefont {Olmschenk}\ \emph {et~al.}(2007)\citenamefont
  {Olmschenk}, \citenamefont {Younge}, \citenamefont {Moehring}, \citenamefont
  {Matsukevich}, \citenamefont {Maunz},\ and\ \citenamefont
  {Monroe}}]{SOlmschenk2007}%
  \BibitemOpen
  \bibfield  {author} {\bibinfo {author} {\bibfnamefont {S.}~\bibnamefont
  {Olmschenk}}, \bibinfo {author} {\bibfnamefont {K.~C.}\ \bibnamefont
  {Younge}}, \bibinfo {author} {\bibfnamefont {D.~L.}\ \bibnamefont
  {Moehring}}, \bibinfo {author} {\bibfnamefont {D.~N.}\ \bibnamefont
  {Matsukevich}}, \bibinfo {author} {\bibfnamefont {P.}~\bibnamefont {Maunz}},
  \ and\ \bibinfo {author} {\bibfnamefont {C.}~\bibnamefont {Monroe}},\ }\href
  {\doibase 10.1103/PhysRevA.76.052314} {\bibfield  {journal} {\bibinfo
  {journal} {Phys. Rev. A}\ }\textbf {\bibinfo {volume} {76}},\ \bibinfo
  {pages} {052314} (\bibinfo {year} {2007})}\BibitemShut {NoStop}%
\bibitem [{\citenamefont {Boldin}\ \emph {et~al.}(2018)\citenamefont {Boldin},
  \citenamefont {Kraft},\ and\ \citenamefont {Wunderlich}}]{IBoldin2018}%
  \BibitemOpen
  \bibfield  {author} {\bibinfo {author} {\bibfnamefont {I.~A.}\ \bibnamefont
  {Boldin}}, \bibinfo {author} {\bibfnamefont {A.}~\bibnamefont {Kraft}}, \
  and\ \bibinfo {author} {\bibfnamefont {C.}~\bibnamefont {Wunderlich}},\
  }\href {\doibase 10.1103/PhysRevLett.120.023201} {\bibfield  {journal}
  {\bibinfo  {journal} {Phys. Rev. Lett.}\ }\textbf {\bibinfo {volume} {120}},\
  \bibinfo {pages} {023201} (\bibinfo {year} {2018})}\BibitemShut {NoStop}%
\bibitem [{\citenamefont {Tycko}(1983)}]{RTycko83}%
  \BibitemOpen
  \bibfield  {author} {\bibinfo {author} {\bibfnamefont {R.}~\bibnamefont
  {Tycko}},\ }\href {\doibase 10.1103/PhysRevLett.51.775} {\bibfield  {journal}
  {\bibinfo  {journal} {Phys. Rev. Lett.}\ }\textbf {\bibinfo {volume} {51}},\
  \bibinfo {pages} {775} (\bibinfo {year} {1983})}\BibitemShut {NoStop}%
\bibitem [{\citenamefont {Barrett}\ \emph {et~al.}(2003)\citenamefont
  {Barrett}, \citenamefont {DeMarco}, \citenamefont {Schaetz}, \citenamefont
  {Meyer}, \citenamefont {Leibfried}, \citenamefont {Britton}, \citenamefont
  {Chiaverini}, \citenamefont {Itano}, \citenamefont
  {Jelenkovi\ifmmode~\acute{c}\else \'{c}\fi{}}, \citenamefont {Jost},
  \citenamefont {Langer}, \citenamefont {Rosenband},\ and\ \citenamefont
  {Wineland}}]{symcool1}%
  \BibitemOpen
  \bibfield  {author} {\bibinfo {author} {\bibfnamefont {M.~D.}\ \bibnamefont
  {Barrett}}, \bibinfo {author} {\bibfnamefont {B.}~\bibnamefont {DeMarco}},
  \bibinfo {author} {\bibfnamefont {T.}~\bibnamefont {Schaetz}}, \bibinfo
  {author} {\bibfnamefont {V.}~\bibnamefont {Meyer}}, \bibinfo {author}
  {\bibfnamefont {D.}~\bibnamefont {Leibfried}}, \bibinfo {author}
  {\bibfnamefont {J.}~\bibnamefont {Britton}}, \bibinfo {author} {\bibfnamefont
  {J.}~\bibnamefont {Chiaverini}}, \bibinfo {author} {\bibfnamefont {W.~M.}\
  \bibnamefont {Itano}}, \bibinfo {author} {\bibfnamefont {B.}~\bibnamefont
  {Jelenkovi\ifmmode~\acute{c}\else \'{c}\fi{}}}, \bibinfo {author}
  {\bibfnamefont {J.~D.}\ \bibnamefont {Jost}}, \bibinfo {author}
  {\bibfnamefont {C.}~\bibnamefont {Langer}}, \bibinfo {author} {\bibfnamefont
  {T.}~\bibnamefont {Rosenband}}, \ and\ \bibinfo {author} {\bibfnamefont
  {D.~J.}\ \bibnamefont {Wineland}},\ }\href {\doibase
  10.1103/PhysRevA.68.042302} {\bibfield  {journal} {\bibinfo  {journal} {Phys.
  Rev. A}\ }\textbf {\bibinfo {volume} {68}},\ \bibinfo {pages} {042302}
  (\bibinfo {year} {2003})}\BibitemShut {NoStop}%
\bibitem [{\citenamefont {Home}\ \emph {et~al.}(2009)\citenamefont {Home},
  \citenamefont {Hanneke}, \citenamefont {Jost}, \citenamefont {Amini},
  \citenamefont {Leibfried},\ and\ \citenamefont {Wineland}}]{symcool2}%
  \BibitemOpen
  \bibfield  {author} {\bibinfo {author} {\bibfnamefont {J.~P.}\ \bibnamefont
  {Home}}, \bibinfo {author} {\bibfnamefont {D.}~\bibnamefont {Hanneke}},
  \bibinfo {author} {\bibfnamefont {J.~D.}\ \bibnamefont {Jost}}, \bibinfo
  {author} {\bibfnamefont {J.~M.}\ \bibnamefont {Amini}}, \bibinfo {author}
  {\bibfnamefont {D.}~\bibnamefont {Leibfried}}, \ and\ \bibinfo {author}
  {\bibfnamefont {D.~J.}\ \bibnamefont {Wineland}},\ }\href {\doibase
  10.1126/science.1177077} {\bibfield  {journal} {\bibinfo  {journal}
  {Science}\ }\textbf {\bibinfo {volume} {325}},\ \bibinfo {pages} {1227}
  (\bibinfo {year} {2009})},\ \Eprint
  {http://arxiv.org/abs/https://science.sciencemag.org/content/325/5945/1227.full.pdf}
  {https://science.sciencemag.org/content/325/5945/1227.full.pdf} \BibitemShut
  {NoStop}%
\bibitem [{\citenamefont {Shen}\ and\ \citenamefont {Lin}(2020)}]{YCShen20}%
  \BibitemOpen
  \bibfield  {author} {\bibinfo {author} {\bibfnamefont {Y.-C.}\ \bibnamefont
  {Shen}}\ and\ \bibinfo {author} {\bibfnamefont {G.-D.}\ \bibnamefont {Lin}},\
  }\href {\doibase 10.1088/1367-2630/ab84b6} {\bibfield  {journal} {\bibinfo
  {journal} {New Journal of Physics}\ }\textbf {\bibinfo {volume} {22}},\
  \bibinfo {pages} {053032} (\bibinfo {year} {2020})}\BibitemShut {NoStop}%
\end{thebibliography}%

\appendix

\section{Up to the second order expansion in $\hat{p}_1$}
\label{sec:app_p1}

Inspecting the $A$ and $B$ functions in (\ref{eq:ABfunc}), 
we notice that $B_0^\pm$ are already in the form of a simple power series 
of $\hat{p}_1$. For the rest, we transform the summations in (\ref{eq:ABfunc})
over $n$ and, if necessary, $m$ into functions in compact
forms with respect to the zeroth, first, and second order
terms of $\hat{p}_1$. All orders in $\hat{q}_1$ are kept.
To simplify the expressions, we define $s_0 = 1/\sqrt{1+p_0^2}$
and $\hat{s}_{\pm} = s_0\sqrt{\pm i(p_0 + \hat{p}_1)}$.
We then have
\begin{widetext}
\begin{align}
    A_1 =& s_0^{1/2}\left[
      1 - \frac{p_0}{2} s_0^2 \hat{p}_1 
      + \frac{3p_0^2-2}{8} s_0^4 \hat{p}_1^2 + \mathcal{O}(\hat{p}_1^3)
    \right] \;, \nonumber \\
%%%%%%%%%%%%%%%%%%%%%%%%%%%%%%%%%%%%%%%
    A_2 =&
      \sum_{l_q=0}^{\infty}(s_0 \hat{q}_1)^{l_q}
      \sum_{n=\lceil l_q/2 \rceil}^{\infty}
      \frac{(-1)^{n}}{n!}\binom{2n}{l_q}(s_0 q_0)^{2n-l_q}
      \left[ 1 - 2 n s_0^2 p_0 \hat{p}_1
      + \{(2 n^2 + n) p_0^2 - n\} s_0^4 \hat{p}_1^2
      + \mathcal{O}(\hat{p}_1^3)
      \right] \nonumber \\
    =& e^{-s_0^2 q_0^2}\sum_{l_q=0}^{\infty}
      \frac{(-s_0 \hat{q}_1)^{l_q}}{{l_q!}}\left[\vphantom{\frac{1^2}{2}}
        \mathcal{H}_{l_q}(s_0 q_0)- s_0^2 p_0\left\{
          l_q\mathcal{H}_{l_q}(s_0 q_0) - s_0 q_0
          \mathcal{H}_{l_q+1}(s_0 q_0)
        \right\} \hat{p}_1 + \right. \nonumber \\
    &\left. s_0^4\left\{
      \left[\frac{p_0^2}{2}(l_q - 2 s_0^2 q_0^2)(l_q+1)-
      \frac{l_q}{2}\right]\mathcal{H}_{l_q}(s_0 q_0)
      +\left[p_0^2 s_0 q_0(s_0^2 q_0^2 - l_q - 1) 
      + \frac{s_0 q_0}{2}\right]\mathcal{H}_{l_q+1}(s_0 q_0)
      \right\} \hat{p}_1^2 + \mathcal{O}(\hat{p}_1^3)
      \vphantom{\frac12}\right]
    \;, \nonumber \\
%%%%%%%%%%%%%%%%%%%%%%%%%%%%%%%%%%%%%%%
    B_1^{\pm} =&
    e^{\pm \frac{i}{2} \arctan (p_0)} \left[
      1 + \frac{\pm i}{2} s_0^2 \hat{p}_1
      - \frac{1 \pm 4 i p_0}{8} s_0^4 \hat{p}_1^2 + \mathcal{O}(\hat{p}_1^3)
    \right]
    \;, \nonumber \\
%%%%%%%%%%%%%%%%%%%%%%%%%%%%%%%%%%%%%%%
    % B_2^{\pm} =&
    %   \sum_{l_q=0}^{\infty}\hat{q}_1^{l_q}
    %   \sum_{n=\lceil l_q/2 \rceil}^{\infty}
    %   \frac{(-1)^{n}s_\pm^{2n}}{n!}\binom{2n}{l_q}q_0^{2n-l_q} 
    %   \left[ 1 + \frac{n (1-p_0^2)}{p_0} s_0^2 \hat{p}_1
    %   + \frac{n\left\{n-1-(2n+4)p_0^2+(n+1)p_0^4\right\}}
    %   {2p_0^2} s_0^4 \hat{p}_1^2 
    %   + \mathcal{O}(\hat{p}_1^3)
    %   \right] \nonumber \\
    % =& e^{-s_\pm^2 q_0^2}\sum_{l_q=0}^{\infty}
    %   \frac{(-1)^{l_q}}{{l_q!}}
    %   \left(s_\pm \hat{q}_1\right)^{l_q}\left[
    %     \mathcal{H}_{l_q}(s_\pm q_0) + 
    %     \frac{1-p_0^2}{2 p_0} s_0^2\left\{
    %       l_q\mathcal{H}_{l_q}(s_\pm q_0) - s_\pm q_0
    %       \mathcal{H}_{l_q+1}(s_\pm q_0)
    %     \right\} \hat{p}_1 + \right. \nonumber \\
    % &\left. \frac{s_0^4}{2p_0^2}\left\{
    %   \left[\frac{l_q(l_q-2)}{4}-
    %   (4l_q+l_q^2)\frac{p_0^2}{2}+(2l_q+l_q^2)\frac{p_0^4}{4}-
    %   (1+l_q)(1-2p_0^2+p_0^4)\frac{s_\pm^2 q_0^2}{2}
    %   \right]\mathcal{H}_{l_q}(s_\pm q_0) +
    % \right.\right. \nonumber \\
    % &\left.\left.
    %   \left[(1-2l_q+10p_0^2+4l_q p_0^2-3p_0^4-2l_q p_0^4)
    %   \frac{s_\pm q_0}{4} + (1-2p_0^2+p_0^4)
    %   \frac{s_\pm^3 q_0^3}{2}
    %   \right]\mathcal{H}_{l_q+1}(s_\pm q_0)
    %   \right\} \hat{p}_1^2 + \mathcal{O}(\hat{p}_1^3)
    %   \vphantom{\frac12}\right] 
    % \;,
    B_2^{\pm} =&
      \sum_{l_q=0}^{\infty}(s_0 \hat{q}_1\sqrt{p_0+\hat{p}_1})^{l_q}
      \sum_{n=\lceil l_q/2 \rceil}^{\infty}
      \frac{(\mp i)^n}{n!}\binom{2n}{l_q}
      (s_0 q_0\sqrt{p_0+\hat{p}_1})^{2n-l_q}
      \left[ 1 - 2 n s_0^2 p_0 \hat{p}_1
      + \{(2 n^2 + n) p_0^2 - n\} s_0^4 \hat{p}_1^2
      + \mathcal{O}(\hat{p}_1^3)
      \right] \nonumber \\
    =& e^{\mp i s_0^2 q_0^2 p_0} B_0^\pm(s_0^2 q_0^2 \hat{p_1})
      \sum_{l_q=0}^{\infty}\frac{(-\hat{s}_\pm \hat{q}_1)^{l_q}}{{l_q!}}
      \left[\vphantom{\frac{1^2}{2}}\mathcal{H}_{l_q}(\hat{s}_{\pm}q_0)
      -s_0^2 p_0\left\{l_q\mathcal{H}_{l_q}(\hat{s}_{\pm}q_0)-
          \hat{s}_{\pm}q_0\mathcal{H}_{l_q+1}(\hat{s}_{\pm}q_0)
        \right\} \hat{p}_1 + \right. \nonumber \\
    &\left. s_0^4\left\{
      \left[\frac{p_0^2}{2}(l_q - 2 \hat{s}_{\pm}^2q_0^2)(l_q+1)-
      \frac{l_q}{2}\right]\mathcal{H}_{l_q}(\hat{s}_{\pm}q_0)
      +\left[p_0^2 \hat{s}_{\pm}q_0(\hat{s}_{\pm}^2q_0^2 - l_q - 1) 
      + \frac{\hat{s}_{\pm}q_0}{2}\right]\mathcal{H}_{l_q+1}(\hat{s}_{\pm}q_0)
      \right\} \hat{p}_1^2 + \mathcal{O}(\hat{p}_1^3)
      \vphantom{\frac12}\right]
    \;,
\label{eq:ABalternative}
\end{align}
\end{widetext}
where $\mathcal{H}_n(x)$ denotes the $n$th-order Hermite polynomial.
The $B_2^{\pm}$ functions are the most complicated, expanded in a power 
series of $\hat{s}_\pm\hat{q}_1$. Despite the square root in the operator
$\hat{s}_\pm$, the total power on each $\hat{s}_\pm$ term is guaranteed
to be an even number, which removes the square root and only leaves integer
powers of the ladder operators. Note that for a specific sum index $l_q$, 
the lowest power of $\hat{q}_1$ that appear in the summand of $B_2^{\pm}$
is $l_q$. Similarly, for a specific $l_q$, the lowest power of $\hat{p}_1$
that appear in the same summand is $2\lceil l_q/2 \rceil$.
% \ming{IMO albeit we eventually basically truncate everything 
% to the zeroth order, leaving the series to the second order here can help show 
% convergence in the series, and the first order will serve as a way to evaluate 
% the contribution of the higher order terms to justify the truncation later.
% Maybe it's good to mention the reasoning here?}
% \nam{yes, please mention why we're doing this explicitly so as to convince
% the readers these expressions are important!}

\end{document}